\documentclass[preprint,notoc]{JHEP3}
\usepackage{epsfig}

\def\eslt{\not\!\!{E_T}}
\def\eslt{E_T^{\rm miss}}

\def\to{\rightarrow}

\def\bi{\begin{itemize}}
\def\ei{\end{itemize}}
\def\te{\tilde e}
\def\DRbar{\overline{DR}}
\def\ta{\tilde a}
\def\tG{\tilde G}

\def\tu{\tilde u}
\def\tc{\tilde c}

\def\tb{\tilde b}

\def\td{\tilde d}

\def\tst{\tilde t}

\def\tg{\tilde g}

\def\tq{\tilde q}

\def\tw{\tilde\chi^\pm}
\def\tz{\tilde\chi^0}
\def\be{\begin{equation}}  
\def\ee{\end{equation}}  
\def\bea{\begin{eqnarray}}  
\def\eea{\end{eqnarray}}

\newcommand{\gsim}{\;\raisebox{-0.9ex}
           {$\textstyle\stackrel{\textstyle >}{\sim}$}\;}
\newcommand{\lsim}{\;\raisebox{-0.9ex}{$\textstyle\stackrel{\textstyle<}
           {\sim}$}\;}
\def\alt{\lsim}
\def\agt{\gsim}

\title{Dark matter allowed scenarios \\ for Yukawa-unified $SO(10)$ SUSY GUTs}

\author{Howard Baer$^a$, Sabine Kraml$^b$, Sezen Sekmen$^c$ and Heaya Summy$^a$\\
$^a$\,Department of Physics, Florida State University Tallahassee, 
FL 32306, USA\\
$^b$\,Laboratoire de Physique de Subatomique et de Cosmologie, 
UJF Grenoble 1, CNRS/IN2P3, INPG, 53 Avenue des Martyrs,
F-38026 Grenoble, France\\
$^c$\,Dept. of Physics, Middle East Technical Univ., TR-06531 Ankara, Turkey\\
E-mail: \email{baer@hep.fsu.edu}, \email{sabine.kraml@cern.ch},
            \email{sezen.sekmen@cern.ch}, \email{heaya@hep.fsu.edu}, 
}

\preprint{\vbox{FSU-HEP-071225, LPSC 07-195}}

\abstract{
Simple supersymmetric grand unified models based on the gauge group
$SO(10)$ require --in addition to gauge and matter unification-- 
the unification of $t$--$b$--$\tau$ Yukawa couplings.
Owing to sparticle contributions to fermion self-energy
diagrams, the Yukawa unification however only occurs for very
special values of the soft SUSY breaking parameters. 
We perform a search using a Markov Chain Monte Carlo (MCMC) technique 
to investigate model parameters and sparticle mass spectra which occur
in Yukawa-unified SUSY models, where we also require the
relic density of neutralino dark matter to saturate the WMAP-measured
abundance.
For Yukawa unified models with $\mu >0$, 
the spectrum is characterizd by three mass scales:
first and second generation scalars in the multi-TeV range, third
generation scalars in the TeV range, and gauginos in the $\sim 100$ 
GeV range. 
Most solutions give far too high a relic abundance of neutralino dark matter.
The dark matter discrepancy can be rectified by 
{\it i}). allowing for  neutralino decay to axino plus photon, 
{\it ii}). imposing gaugino mass non-universality
or {\it iii}). imposing generational non-universality.
In addition, the MCMC approach finds a compromise solution where 
scalar masses are not too heavy, 
and where neutralino annihilation occurs via the light Higgs $h$ resonance.
By imposing weak scale Higgs soft term boundary conditions, 
we are also able to generate low $\mu,\ m_A$ solutions with 
neutralino annihilation via a light $A$ resonance, though these
solutions seem to be excluded by CDF/D0
measurements of the $B_s\to \mu^+\mu^- $ branching fraction.
Based on the dual requirements of Yukawa coupling unification and 
dark matter relic density, we predict new physics signals at the LHC from 
pair production of 350--450 GeV gluinos. The events are characterized by 
very high $b$-jet multiplicity and a dilepton mass edge around 
$m_{\tz_2}-m_{\tz_1}\sim 50$--$75$ GeV.
}

\begin{document}

\tableofcontents

\section{Introduction}
\label{sec:intro}

Grand unified theories (GUTs) based upon the gauge group $SO(10)$
certainly have 
to be considered as among the most beautiful ideas in particle physics~\cite{so10}.
In addition to gauge group unification, one also has matter unification 
of each generation within the $SO(10)$ 16-dimensional spinorial
representation $\psi (16)$. Furthermore, the simplest $SO(10)$ GUT theories
also allow for Yukawa coupling unification. The ad-hoc but fortuitous
triangle anomaly cancellation found in the SM or even $SU(5)$  GUTs
is a simple mathematical fact in $SO(10)$.
The beauty of $SO(10)$ is only enhanced via a marriage to softly broken 
$N=1$ supersymmetry (SUSY), which stabilizes the  gauge hierarchy, and is experimentally 
supported by gauge coupling unification found within the MSSM 
(provided superpartners exist at or around the weak scale).
SUSY $SO(10)$ also elegantly addresses the neutrino mass problem, since one only has
matter unification within the superfield $\hat{\psi}(16)$ provided one introduces an 
additional (SM gauge singlet) superfield $\hat{N}^c$ which contains a 
right-handed neutrino state. Under the breaking of $SO(10)$, a superpotential term
leading to a Majorana mass is allowed for the $\hat{N}^c$ field; this
naturally yields a description of neutrino masses in terms of the 
elegant see-saw mechanism~\cite{seesaw}.

Standard GUTs and also SUSYGUTs formulated in 4-d spacetime
have fallen into dis-repute due to a variety of problems 
associated with GUT gauge symmetry breaking via the Higgs mechanism.
These include the doublet-triplet splitting problem, 
lack of observation of proton decay, and the frequently awkward
implementation of GUT symmetry breaking via at least one large and unwieldy 
Higgs representation. With the onset of model building utilizing extra
dimensions,  it has been shown to be possible to formulate SUSYGUTs
in 5 or more spacetime dimensions. Then, the GUT gauge symmetry can be broken 
via compactification of the extra dimensions on a suitable sub-space, such as 
an orbifold. In these 5-d and 6-d SUSYGUT models, the large GUT scale Higgs 
representations can be dispensed with, the doublet-triplet splitting problem can be 
solved, and the proton can be made longer-lived than current limits or even absolutely 
stable~\cite{exdimguts}. The extra-dimensional SUSYGUT models act as a sort of ``proof of 
principle'' of what might be possible in more complicated set-ups where
the SUSYGUT model might arise from compactification of superstring models.

The imminent turn-on of the CERN LHC naturally leads one to ask: How might $SO(10)$
SUSYGUT theories manifest themselves in the environment of an 
LHC detector?
Our goal in this paper is to address this question. To do so, our path will be guided
by the twin requirements of {\it i}). Yukawa coupling unification and 
{\it ii}). explaining the measured dark matter (DM) abundance of the universe.
Our answer we find is that, if DM-allowed, Yukawa-unified $SO(10)$ SUSYGUTs 
are correct, then we expect new physics at LHC to consist of gluino pair production events
with $m_{\tg}\sim 350$--$450$ GeV, followed by 3-body gluino cascade decays to 
$b$-jet rich final states. In addition, an opposite sign/same-flavor (OS/SF) isolated
dilepton invariant mass spectrum should have a mass edge at around $50$--$75$ GeV.
SUSY scalar fields other than the light Higgs $h$ are very heavy, and largely 
decouple from LHC physics.
The remainder of this paper details our methodology as to how we come to these
conclusions.

In this paper, we assume nature is described by an $SO(10)$ SUSYGUT theory
at energy scales $Q>M_{GUT}\sim 2\times 10^{16}$ GeV.
We further assume that the $SO(10)$ SUSYGUT model breaks 
(either via the Higgs mechanism or via compactification of extra dimensions) 
to the MSSM (or MSSM plus right-handed neutrino states) at $Q=M_{GUT}$. 
Thus, below $M_{GUT}$, the MSSM is the correct effective field theory 
which describes nature. 
We will further assume that the superpotential
above $M_{GUT}$ is of the form
\be
\hat{f}\ni f\hat{\psi}_{16}\hat{\psi}_{16}\hat{\phi}_{10} +\cdots
\ee 
so that the three third generation Yukawa couplings $f_t$, $f_b$ and $f_\tau$ are
unified at $M_{GUT}$. It is simple in this context to include as well the effect of
a third generation neutrino Yukawa coupling $f_\nu$; this effect has been shown
to be small, although it can help improve Yukawa coupling unification 
by a few per cent if the 
neutrino Majorana mass scale is within a few orders of magnitude of $M_{GUT}$.
Within this ansatz, the GUT scale soft SUSY breaking (SSB) 
terms are constrained by the $SO(10)$ gauge symmetry so that matter scalar 
SSB terms have a 
common mass $m_{16}$, Higgs scalar SSB terms have a common mass $m_{10}$ and there is a common 
trilinear soft breaking parameter $A_0$.
As usual, the bilinear soft term $B$ 
can be traded for $\tan\beta$, the ratio of Higgs 
field vevs, while the magnitude of the superpotential Higgs mass $\mu $ is 
determined in terms of $M_Z^2$ via the electroweak symmetry breaking 
minimization conditions. Here, electroweak symmetry is broken radiatively (REWSB) 
due to the large top quark mass.
In order to accomodate REWSB, it is well-known that in Yukawa-unified models, 
the GUT scale Higgs soft masses must be
{\it split} such that $m_{H_u}^2 <m_{H_d}^2$ in order to fulfill the EWSB minimization 
conditions; this effectively gives $m_{H_u}^2$ a head start over $m_{H_d}^2$
in running towards negative values at or around the weak scale.
We parametrize the Higgs splitting as $m_{H_{u,d}}^2=m_{10}^2\mp 2M_D^2$. 
The Higgs mass splitting might  originate via a large near-GUT-scale threshold correction 
arising from the neutrino Yukawa coupling: see the Appendix to Ref.~\cite{bdr} for discussion. 
Thus, the Yukawa unified SUSY model is determined by the parameter space
\be
m_{16},\ m_{10},\ M_D^2,\ m_{1/2},\ A_0,\ \tan\beta, \ sign(\mu )
\label{eq:pspace}
\ee
along with the top quark mass. We will take $m_t=171$ GeV, in accord with recent 
measurements from CDF and D0~\cite{mtop}.

Much previous work has been done on the topic of $t$--$b$--$\tau$ 
unification in $SO(10)$ SUSYGUTs.
Early on it was found that $t$--$b$--$\tau$ Yukawa unification could only occur at very high 
values of $\tan\beta$\cite{old}. 
The importance of weak scale MSSM threshold corrections
to fermion masses was noted by Hall, Rattazzi and Sarid\cite{hrs}. It was also noted that
with Yukawa coupling unification, in order to obtain an appropriate REWSB, 
the GUT scale Higgs masses would need to be split such that 
$m_{H_u}^2<m_{H_d}^2$-- perhaps via $D$-term contributions\cite{dterms} to {\it all} 
scalar masses (the DT model), 
or via splitting of {\it only} the Higgs soft terms\cite{bdr} (the HS model).

In Ref.~\cite{bdft}, it was found using the Isajet sparticle mass 
spectrum generator~\cite{isajet} Isasuga that
Yukawa coupling unification to 5\% could be achieved in the MSSM using $D$-term
splitting, but only for $\mu <0$; for $\mu >0$, the Yukawa coupling unification 
was much worse, of order 30--50\%. These parameter space scans allowed $m_{16}$
values up to only 1.5 TeV, and used a GUT scale Yukawa unification quantity 
\be
R=\frac{max(f_t,f_b,f_\tau )}{min(f_t,f_b,f_\tau )},
\label{eq:R}
\ee
so that {\it e.g.} $R=1.1$ would correspond to 10\% Yukawa unification.
The $\mu <0$ Yukawa unification solutions were examined in more detail
in Ref.~\cite{bbdfmqt}, where dark matter allowed solutions were found,
and the neutralino $A$-annihilation funnel was displayed for the first time.

With the announcement from BNL experiment E-821 that there was a $3\sigma$
deviation from SM predictions on the muon anomalous magnetic moment 
$a_\mu\equiv (g-2)_\mu /2$, attention shifted back to $\mu >0$
solutions. Ref.~\cite{bf}, using the DT model with parameter
space scans of $m_{16}$ up to 2 TeV, found Yukawa-unified solutions with $R\sim 1.3$
but only for special choices of GUT scale boundary conditions:
\be
A_0\sim -2\,m_{16},\ m_{10}\sim 1.2\,m_{16},
\label{eq:BCs}
\ee
with $m_{1/2}\ll m_{16}$ and $\tan\beta \sim 50$.
In fact, these boundary conditions had been found earlier by
Bagger {\it et al.}~\cite{bfpz} in the context of models with a 
{\it radiatively driven inverted scalar mass hierarchy} (RIMH), wherein
RG running of multi-TeV GUT scale scalar masses caused third generation
masses to be driven to weak scale values, while first/second generation soft terms remained
in the multi-TeV regime. These models, which required
Yukawa coupling unification, were designed to maintain low fine-tuning  by having
light third generation scalars, while solving the SUSY flavor and CP problems
via multi-TeV first and second generation scalars. A realistic implementation
of these models in Ref.~\cite{imh} using 2-loop RGEs and requiring REWSB found
that an inverted hierarchy could be generated, but only to a lesser extent
than that envisioned in Ref.~\cite{bfpz}, which didn't implement EWSB or calculate
an actual physical mass spectrum. 

Simultaneously with Ref.~\cite{bf}, Blazek, Dermisek and Raby (BDR) published 
results showing Yukawa-unified solutions using the HS model solution~\cite{bdr}.
Their results {\it also} found valid solutions using the Bagger {\it et al.}
boundary conditions. BDR used a top-down method beginning with
actual Yukawa unification at $M_{GUT}$, and implemented 2-loop gauge and Yukawa
running but 1-loop soft term running. 
They extracted physical soft terms at 
scale $Q=M_Z$, and minimized a two-Higgs doublet scalar potential to 
achieve REWSB, also at scale $M_Z$. Each run generated a numerical value for 
third generation $t$, $b$ and $\tau$ masses and other electroweak and QCD observables.
A $\chi^2$ fit was performed to select those solutions which best matched the
meaured weak scale fermion masses and other parameters.
BDR scanned $m_{16}$ values up to 2 TeV, and found best fit results with $m_A\sim 100$ GeV
and $\mu\sim 100$--$200$ GeV,
in contrast to Ref.~\cite{bf}, where 
solutions with valid EWSB could only be found if $\mu\sim m_A\sim m_{\tst_1}\sim 1$ TeV.\footnote{
A paper by Tobe and Wells\cite{tw} (TW) appeared after Ref. \cite{bdr}. While TW calculate no
actual spectra or address EWSB, they do adopt a semi-model independent approach which
favors $t-b-\tau$ Yukawa coupling unification if scalar masses are in the multi-TeV regime while
gauginos are as light as possible.} 

In a long follow-up study using Isajet, Auto {\it et al.}~\cite{abbbft} found that 
Yukawa-unified solutions good to less than a few percent could be found in the $\mu >0$ case 
using the HS model of BDR, but only for very
large values of $m_{16}\agt 5$--$10$ TeV and low values of $m_{1/2}\alt 100$ GeV, 
again using Bagger {\it et al.} boundary conditions. 
Yukawa unification in the DT model was at best good to 10\% (for this reason, in the present paper
we will focus only on the HS model).
The spectra were characterized by three mass scales:
\begin{enumerate}
\item $\sim 5$--$15$ TeV first and second generation scalars, 
\item  $\sim 1$ TeV third generations scalars, $\mu$ term and $m_A$ and 
\item  chargino masses $m_{\tw_1}\sim 100$--$200$ GeV and gluino masses
$m_{\tg}\sim 350$--$450$ GeV. 
\end{enumerate}
These Yukawa-unified solutions--
owing to very large values of scalar masses, $m_A$ and $\mu$-- predicted dark matter
relic density values $\Omega_{\tz_1}h^2$ far beyond the WMAP-measured result~\cite{wmap} of 
\be
\Omega_{\rm DM}h^2 = 0.111^{+0.011}_{-0.015} \ \ (2\sigma).
\label{eq:wmap}
\ee
Meanwhile, the spectra generated using the BDR
program could easily generate $\Omega_{\tz_1}h^2$ values close to 0.1 since their
allowed $\mu$ and $m_A$ values were far lower, so that mixed higgsino dark matter or
$A$-funnel annihilation solutions could easily be found. In follow-up papers
to the BDR program~\cite{drrr1,drrr2}, the neutralino relic density and branching fraction
$B_s\to\mu^+\mu^- $ were evaluated. To avoid constraints on $BF(B_s\to \mu^+\mu^- )$ from
the CDF collaboration, the best fit values of $m_{16}$ and $m_A$ have been steadily increasing, 
so that the latest papers have $m_A\sim 500$ GeV and $m_{16}\sim 3$ TeV, 
while $\mu$ can still be of order 100 GeV~\cite{drrr2}. 
In Ref.~\cite{auto2}, attempts were made to reconcile the Isajet Yukawa-unified solutions 
with the dark matter relic density. The two solutions advocated were 1.\ lowering 
GUT scale first/second generation scalars relative to the third, 
to gain neutralino-squark or neutralino-slepton co-annihilation solutions~\cite{nmh}, or
2.\ increasing the GUT scale gaugino mass $M_1$, so the relic density could be 
lowered by bino-wino co-annihilation~\cite{bwca}.

In this paper, we report on a new examination of SUSY mass spectra constrained by 
Yukawa coupling unification.
We are motivated by the following rationale.
\begin{itemize}
\item  We use the latest measured value of the top quark mass-- $m_t=171$ GeV~\cite{mtop}--
whereas earlier analyses used $m_t=175$ GeV or even higher values.
\item We generate SUSY mass spectra using Isajet 7.75, which includes several
upgrades to the SUSY spectrum calculation from versions used in earlier analyses~\cite{bfkp}.
\item We use a Markov Chain Monte Carlo (MCMC) technique, which is much more 
efficient at searching through 
multi-dimensional parameter spaces.
\item We adopt an alternative approach to implementing Higgs sector SSB
boundary conditions, which simultaneously imposes GUT-scale Higgs (GSH) and weak-scale
Higgs (WSH) soft term boundary conditions. 
This allows us to explore solutions akin to those found by
BDR with low values of $m_A$ and $\mu$. These solutions are difficult to generate using a purely
top-down approach, due to a high degree of fine-tuning in the EWSB sector.
\end{itemize}

\noindent
In Sections \ref{sec:scenarios} and \ref{sec:mcmc}, we present results from our calculations.
These include:
\begin{enumerate}
\item A variety of solutions consistent with Ref.~\cite{abbbft} with very large
values of $m_{16}\sim 5$--$15$ TeV, and very light gauginos. These solutions
always have {\it too large} a dark matter relic density. 
We propose they may still be valid by invoking an unstable lightest 
neutralino which decays to an {\it axino} lightest SUSY particle (LSP).
\item New solutions are found with $m_{16}\sim 3$ TeV and $m_{1/2}\sim 100$ GeV 
with Yukawa unification to $\sim 8\%$, but with 
a valid relic density due to neutralino annihilation through the light Higgs $h$
resonance.
\item Using simultaneous GSH and WSH boundary conditions, we are able to generate
BDR-like solutions using Isajet with $m_{16}\sim 3$--$7$ TeV but with $m_{1/2}\sim 100$--$400$ GeV
with good relic density due to neutralino annihilation through the
pseudoscalar Higgs $A$ resonance. The $A$-resonance annihilation solutions 
have excellent Yukawa unification, but with $m_A\sim 150$--$230$ GeV and $\tan\beta \sim 50$, 
are all excluded by measurements of the $B_s\to\mu^+\mu^-$ branching fraction.
Solutions with higher $m_A$ and low $\mu$ with mixed higgsino dark matter also
can be found, but tend to have Yukawa unification $R>1.2$.
\item We also re-examine the first class of solutions, but solving the DM problem
via generational non-universality or gaugino mass non-universality,
as in Ref.~\cite{auto2}.
\end{enumerate}

In Sec. \ref{sec:lhc}, we discuss a Table of benchmark scenarios for each of these
cases. The cases are suitable for collider event generation.
We also discuss implications of each of these scenarios
for collider searches at the CERN LHC. 

Our preferred solutions-- numbers 1. and 2. above--
lead to spectra with first/second generation scalar masses in the $3-15$ TeV range,
while gluino masses are in the $350-450$ GeV range. 
(Solution number 3 is likely ruled out by the measured limit on $BF(B_s\to \mu^+\mu^-)$ and 
solution 4 gives up either gaugino mass unification or generational unification.)
Thus, we predict that DM-allowed Yukawa-unified
$SO(10)$ SUSY GUT models will lead to collider events at the CERN LHC typified by
nearly pure gluino pair production followed by cascade decays $\tg\to tb\tw_1$, 
$b\bar{b}\tz_2$ and $b\bar{b}\tz_1$. The gluino pair events will occur 
with cross sections in the $\sim 10^5$ fb range, and will be very rich in $b$-jets.
There should exist a characteristic OS/SF dilepton  mass
edge at $m_{\tz_2}-m_{\tz_1}\sim 50$--$75$~GeV.

A summary and conclusions are presented in Sec. \ref{sec:conclude}.

\section{Random scan in HS model}
\label{sec:scenarios}

In this section, we explore the parameter space Eq.~(\ref{eq:pspace}) for Yukawa-unified 
solutions by means of a random scan. 
We wish to first check and update results presented in Ref.~\cite{abbbft}, using the latest
Isajet version and a top quark mass of $m_t=171$ GeV in accord with recent 
measurements from the Fermilab Tevatron~\cite{mtop}. 
The degree of Yukawa unification, $R$, is defined in  Eq.~(\ref{eq:R}), 
so that {\it e.g.} a value of $R=1.1$ corresponds to 10\% Yukawa unification.

For our calculations, we adopt the Isajet 7.75~\cite{isajet,bfkp} SUSY spectrum generator Isasugra.
Isasugra begins the calculation of the sparticle mass spectrum with
input $\overline{DR}$ gauge couplings and $f_b$, $f_\tau$ Yukawa couplings at the 
scale $Q=M_Z$ ($f_t$ running begins at $Q=m_t$) and evolves the 6 couplings up in energy 
to scale $Q=M_{GUT}$ (defined as the value $Q$ where $g_1=g_2$) using two-loop RGEs.\footnote{
As inputs, we take the top quark pole mass $m_t=171$ GeV. We also take 
$m_b^{\DRbar}(M_Z)=2.83$ GeV\cite{bfmt} and $m_\tau^{\DRbar} (M_Z)=1.7463$ GeV.
The paper Ref. \cite{abbbft} addresses consequences of varying the values
of $m_t$ and $m_b$.}  
At $Q=M_{GUT}$, the SSB boundary conditions are input, and  the set
of 26 coupled two-loop MSSM RGEs~\cite{mv} are used to evolve couplings and SSB terms
 back down in scale to $Q=M_Z$. 
Full two-loop MSSM RGEs are used for soft term evolution, while the gauge and Yukawa coupling
evolution includes threshold effects in the one-loop beta-functions, so the gauge and
Yukawa couplings transition smooothly from the MSSM to SM effective theories as 
different mass thresholds are passed.
In Isajet 7.75, 
the SSB terms of sparticles 
which mix are frozen out at the
scale $Q\equiv M_{SUSY}=\sqrt{m_{\tst_L} m_{\tst_R}}$, while non-mixing SSB terms are frozen out
at their own mass scale~\cite{bfkp}. 
The scalar potential is minimized using the RG-improved one-loop MSSM effective
potential evaluated at an optimized scale $Q=M_{SUSY}$ which accounts for
leading two-loop effects~\cite{haber}.
Once the tree-level sparticle mass spectrum is computed, full one-loop
radiative corrections are caculated for all sparticle and Higgs boson masses,
including complete one-loop weak scale threshold corrections for the
top, bottom and tau masses at scale $Q=M_{SUSY}$. These fermion self-energy 
terms are critical to evaluating whether or not Yukawa couplings do
indeed unify.
Since the GUT scale Yukawa couplings are modified by the threshold corrections, the
Isajet RGE solution must be imposed iteratively with successive up--down 
running until a convergent 
solution for the spectrum is found.
For most of parameter space, there is very good agreement between Isajet and
the other public spectrum codes SoftSusy, SuSpect and SPheno, 
although at the edges of parameter space agreement between the four codes 
typically diminishes~\cite{kraml}.

We first adopt a wide parameter range scan, and then once the best Yukawa-unified regions are
found, we adopt a narrow scan to try to hone in on the best unified solutions.
The parameter range we adopt for the wide (narrow) scan is
\be
\begin{array}{lll}
m_{16}:              & 0\;\textrm{--}\;20\ {\rm TeV}\quad & (1\;\textrm{--}\;20\ {\rm TeV}),\\
m_{10}/m_{16}:\quad & 0\;\textrm{--}\;1.5            & (0.8\;\textrm{--}\;1.4),\\
m_{1/2}:             &  0\;\textrm{--}\;5\ {\rm TeV}     &  (0\;\textrm{--}\;1\ {\rm TeV}),\\
A_0/m_{16}        & -3\;\textrm{--}\; 3                     & (-2.5\;\textrm{--}\;1.9),\\
M_D/m_{16}:      &  0\;\textrm{--}\;0.8                   & (0.25\;\textrm{--}\;0.8),\\
\tan\beta :           & 40\;\textrm{--}\;60                   & (46\;\textrm{--}\;53) .
\end{array}
\label{eq:prange}
\ee
For the random scan, we evaluate $\Omega_{\tz_1}h^2$, $BF(b\to s\gamma )$, $\Delta a_\mu$ and 
$BF(B_S\to \mu^+\mu^- )$ using Isatools (a sub-package of Isajet).
We plot only solutions for which $m_{\tw_1}>103.5$ GeV, in accord with 
LEP2 searches, and for the moment implement no other constraints, such 
as relic density, Higgs mass, {\it etc.}.

\subsection{Random scan results}

Our first results are shown in Fig.~\ref{fig:pspace}, where
we show points from the wide scan (dark blue) and points from 
the narrow scan (light blue) in the parameter versus $R$ plane.
From frame {\it a}), we see that Yukawa unification to better than
30\% ($R<1.3$) cannot be achieved for $m_{16}<1$ TeV, while 
Yukawa coupling unification becomes much more likely at multi-TeV
values of $m_{16}$. Frame {\it b}) shows that Yukawa-unified models
prefer $m_{10}\sim 1-1.3\,m_{16}$, while frame {\it c}) shows that a
positive value of $M_D^2$ $\sim (0.25-0.5)\,m_{16}$-- which yields
$m_{H_u}^2<m_{H_d}^2$-- is preferred. 
In frame {\it d}), we see that the best Yukawa-unified solutions are found for
the lowest possible values of $m_{1/2}$. We note here that-- using 1-loop
RGEs along with the LEP2 constraint $m_{\tw_1}>103.5$ GeV-- one would expect
from models with gaugino mass unification that since $m_{\tw_1}\sim M_2(weak)\sim 0.8 m_{1/2}$ 
that we would have $m_{1/2}\agt 125$ GeV always.
However, the very large values of $m_{16}$ we probe alter the simple 
1-loop gaugino mass unification condition 
(that $\frac{M_1}{\alpha_1}=\frac{M_2}{\alpha_2}=\frac{M_3}{\alpha_3}$) 
via 2-loop RGE effects.
Thus, values of $m_{1/2}$ much lower than $\sim 125$ GeV are possible if $m_{16}$ is large.

In frame {\it e}), we see a sharp dependence that Yukawa-unified solutions 
can only be obtained for $A_0\sim -2m_{16}$, while frame {\it f}) shows
that $\tan\beta$ must indeed be large: in the range $\sim 47-53$.
Bagger {\it et al.} had shown in Ref.~\cite{bfpz} that a radiatively-driven
inverted scalar mass hierarchy with $m(third\ generation)\ll m(first/second\ generation)$
could be derived provided one starts with unified Yukawa couplings, the
boundary conditions 
\be
  4\,m_{16}^2=2\,m_{10}^2=A_0^2
\label{eq:bfpz}
\ee
and one neglects the effect of gaugino masses. Our results in Fig.~\ref{fig:pspace}
show the {\it inverse} effect: that Yukawa coupling unification can only be achieved
if one imposes the boundary conditions (\ref{eq:bfpz}) along with $m_{16}\gg m_{1/2}$.
This result holds only in our numerical calculations for $\mu >0$ and $A_0<0$
and of course $m_{H_u}^2<m_{H_d}^2$.
The results shown in Fig.~\ref{fig:pspace} also verify that the results
obtained in Ref. \cite{abbbft} still hold, even with updated spectra code and a lower
value of $m_t=171$ GeV.


In Fig.~\ref{fig:mass1}, we show various -ino masses\footnote{We collectively refer to the set of all gluinos, 
charginos and neutralinos as -inos.} versus $R$ as generated from our random
scan.
In frame {\it a}), we see that-- owing to the preference of Yukawa-unified solutions to
have $m_{1/2}$ as small as possible, the chargino mass $m_{\tw_1}$ is preferred to be quite light,
as close to the LEP2 limit as possible, with $m_{\tw_1}\sim 100$--$200$ GeV. Likewise, in frame
{\it b}), the gluino mass should be relatively light, with $m_{\tg}\sim 350$--$500$ GeV.
The lightest neutralino $\tz_1$ mass is shown in frame {\it c}), and is preferred
in the range $m_{\tz_1}\sim 50$--$100$ GeV. Meanwhile, the mass difference $m_{\tz_2}-m_{\tz_1}$
is shown in frame {\it d}), and is also in the range $\sim 50$--$100$ GeV. This latter quantity is
important because if $m_{\tz_2}-m_{\tz_1}<M_Z$, two body spoiler decay modes such as
$\tz_2\to \tz_1 Z$ will be kinematically closed, and the three body decays $\tz_2\to\tz_1\ell\bar{\ell}$
($\ell =e\ or\ \mu$) should occur at a sufficiently large rate at the LHC that an edge
should be visible in the $m(\ell\bar{\ell})$ invariant mass distribution at 
$m_{\tz_2}-m_{\tz_1}$~\cite{mlledge}. This measureable mass edge can serve as the starting point for
sparticle mass reconstruction in SUSY particle cascade decay events at the LHC~\cite{frank}.
Thus, in Yukawa-unified models, this mass edge is highly likely to be visible.

In Fig.~\ref{fig:mass2}, we show the expected masses of {\it a}) $\tu_L$-squark, 
{\it b}) the $\tst_1$-squark, {\it c}), the pseudoscalar Higgs boson $A$ and
{\it d}) the superpotential Higgs parameter $\mu$. Frame {\it a}) shows that 
Yukawa-unified solutions prefer first/second generation squarks and sleptons 
with masses in the $5$--$20$ TeV range-- far higher than values typically examined
in phenomenological SUSY studies!
The top squark mass and the $A$, $H$ and $H^\pm$ Higgs bosons 
tends to be somewhat lighter: in the $2$--$8$ TeV range. Finally, frame {\it d})
shows that the $\mu$ parameter-- which is derived from the EWSB minimization conditions--
tends also to be in the $5$--$15$ TeV range. Thus, using a top-down approach to search for
Yukawa-unified solutions in the HS model, we find that $\mu\gg M_1,\ M_2$, so that
the lighter charginos and neutralinos should be gaugino-like, and quite light, while the
heavier charginos and neutralino will be in the muti-TeV range, and nearly pure
higgsino-like states. In particular, the lightest SUSY particle (LSP)-- the neutralino $\tz_1$--
is nearly pure bino-like.

\FIGURE[p]{
\epsfig{file=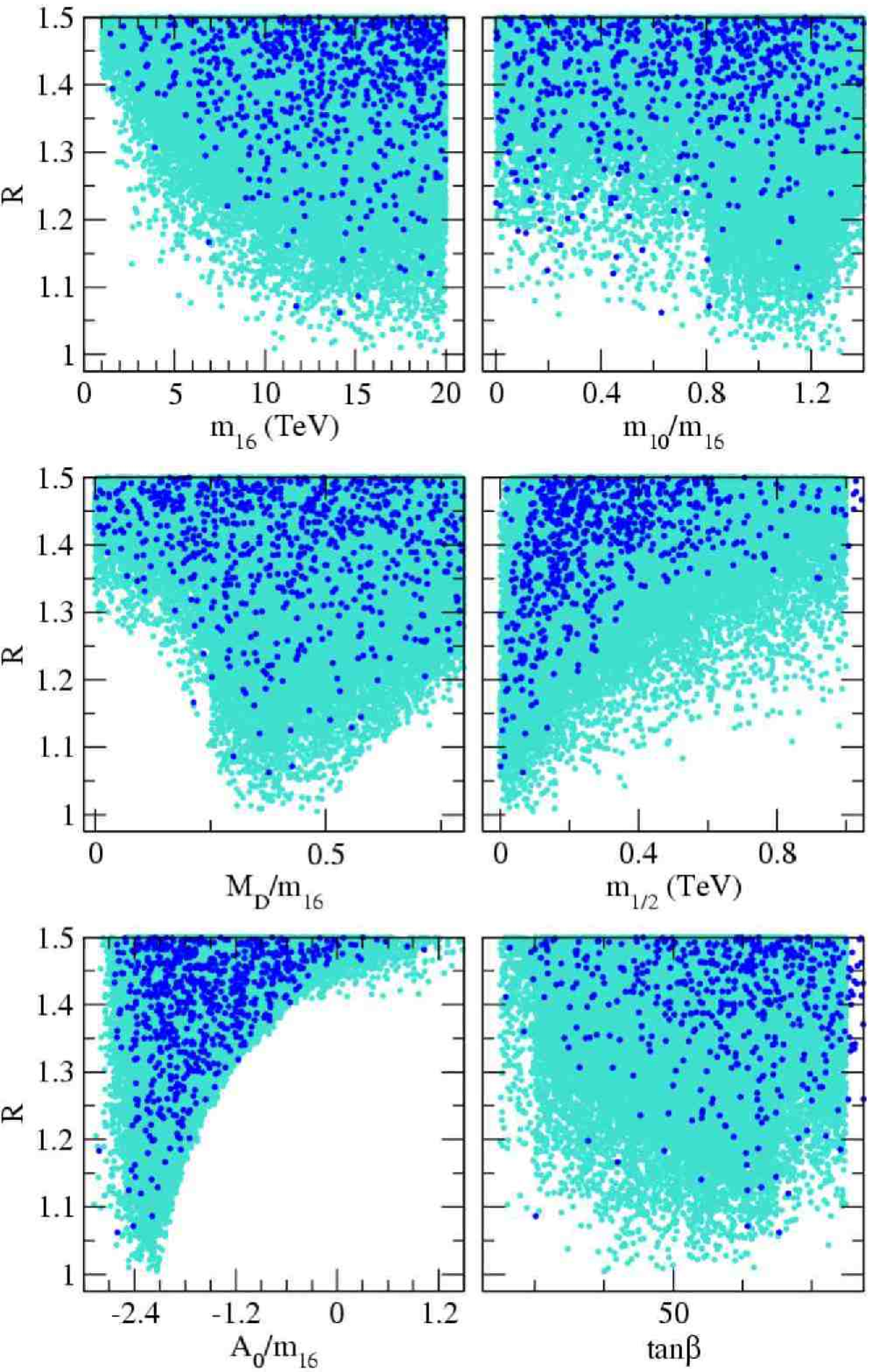,width=12cm} 
\caption{Plot of $R$ versus various input
parameters for a wide (dark blue) and narrow (light blue) 
random scan over the parameter ranges listed in Eq.~(\ref{eq:prange}).
We take $\mu >0$ and $m_t=171$ GeV.
}\label{fig:pspace}}

\FIGURE[p]{
\epsfig{file=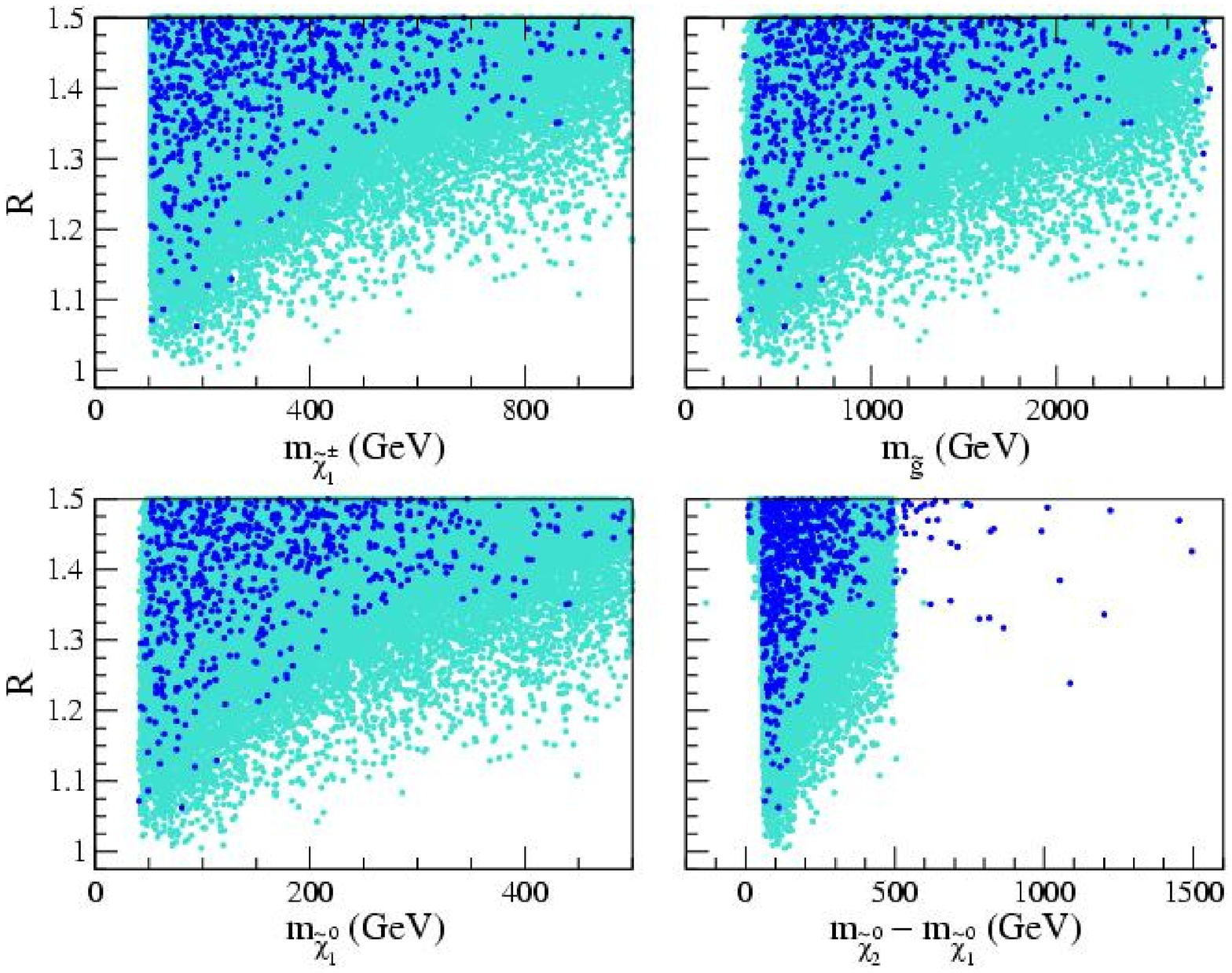,width=12cm} 
\caption{Plot of $R$ versus various sparticle
masses for a random scan over the parameter range listed
in Eq.~(\ref{eq:prange}).
We take $\mu >0$ and $m_t=171$ GeV.
}\label{fig:mass1}}

\FIGURE[p]{
\epsfig{file=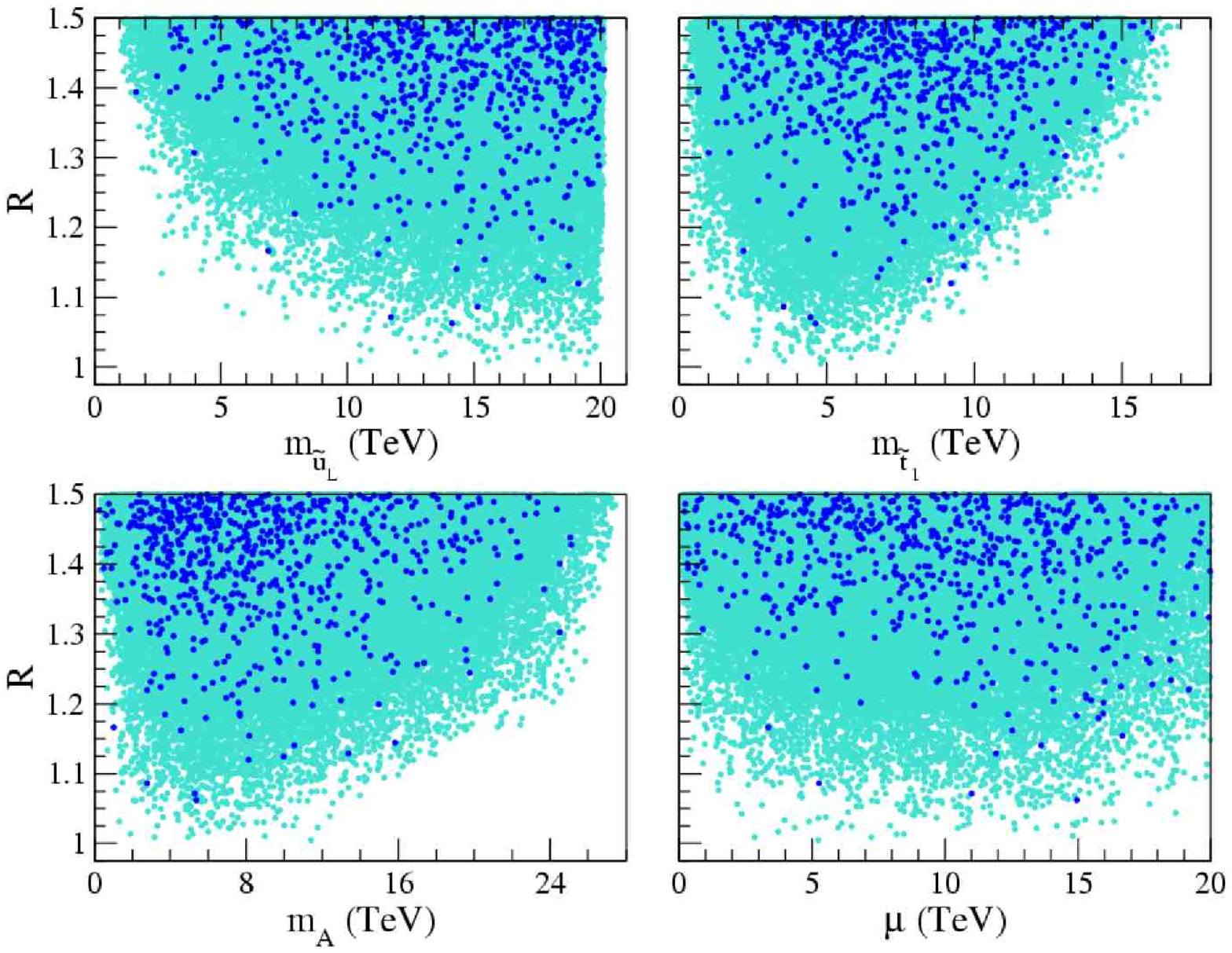,width=12cm} 
\caption{Plot of $R$ versus various sparticle
masses for a random scan over the parameter range listed
in Eq.~(\ref{eq:prange}).
We take $\mu >0$ and $m_t=171$ GeV.
}\label{fig:mass2}}

In Fig.~\ref{fig:Oh2R}, we plot $R$~vs.~$\Omega_{\tz_1}h^2$ for LEP2 allowed points from
our random scan. It is clear that $R\sim 1$ points predict an extremely large 
value of $\Omega_{\tz_1}h^2$ of 30--30,000. On the other hand, if we require
consistency with the WMAP-measured value of $\Omega_{\tz_1}h^2\simeq 0.1$, then
we generate Yukawa-unified solutions to 40\% unification with the random scan.
This plot underscores the difficulty of finding sparticle mass spectra solutions
which are compatible with {\it both} the measured dark matter abundance and $t$--$b$--$\tau$
Yukawa coupling unification.

\clearpage 

\FIGURE[t]{
\epsfig{file=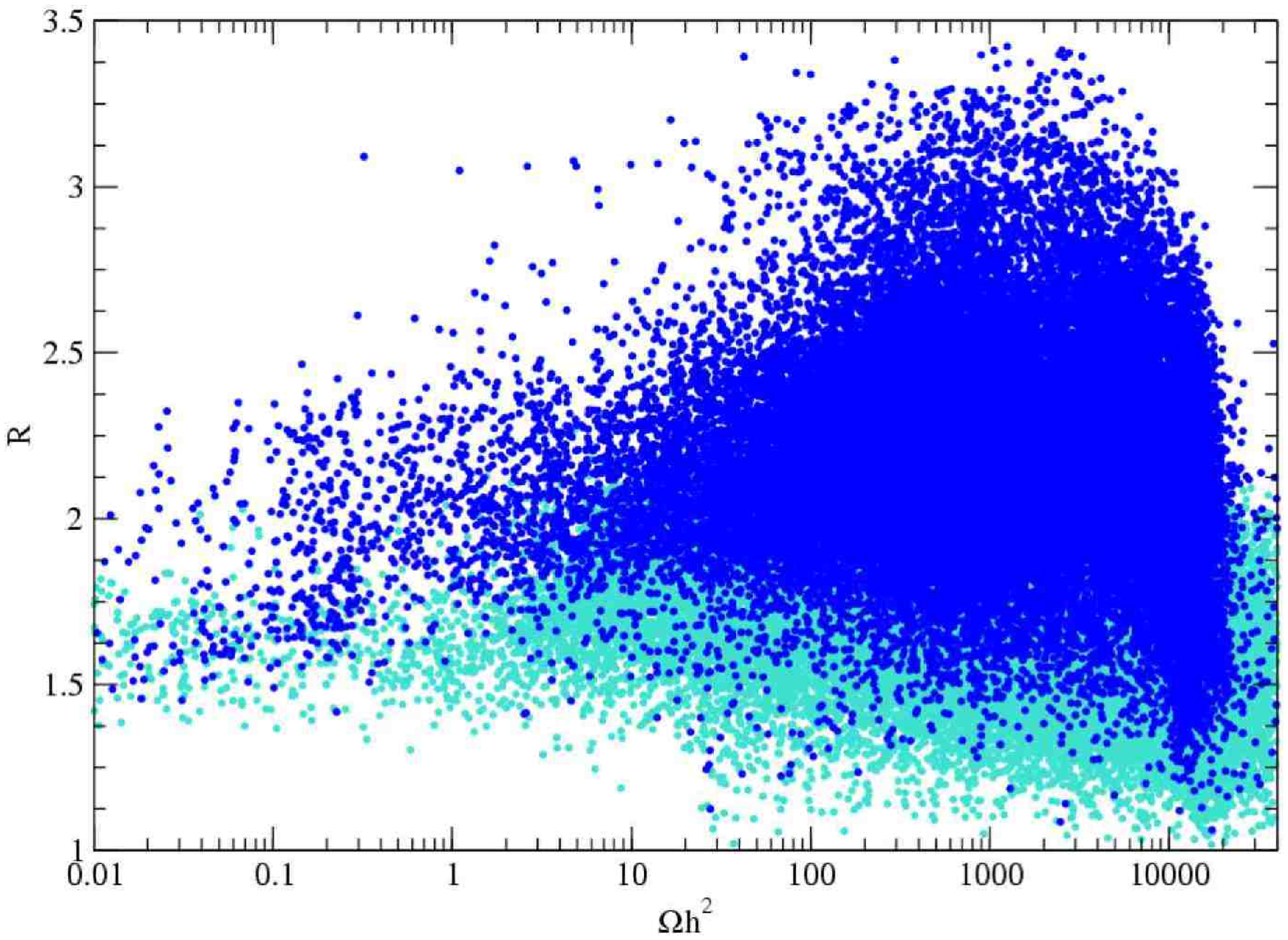,width=10cm} 
\caption{Plot of $\Omega_{\tz_1}h^2$~vs.~$R$
for a random scan over the parameter range listed
in Eq.~(\ref{eq:prange}). We take $\mu >0$ and $m_t=171$ GeV.
}\label{fig:Oh2R}}

\subsection{Three proposals to reconcile Yukawa-unified models with
dark matter relic density}

\subsubsection{Dark matter solution via neutralino decay to axino}

We see from Fig.~\ref{fig:Oh2R} that models generated from the random scan with
$R\sim 1.0$ all have $\Omega_{\tz_1}h^2\sim 30-30,000$: far beyond the WMAP-measured
result of $\Omega_{CDM}h^2\sim 0.1$. One possible solution to reconcile the predicted
and measured dark matter density is to assume that the lightest neutralino $\tz_1$
is in fact not the LSP, but is unstable. Some alternative LSP candidates consist of
the gravitino $\tG$ or the axino $\ta$. In gravity-mediated SUSY breaking models, 
the gravitino mass $m_{3/2}$ arises due to the superHiggs mechanism, and is expected
to set the scale for all the soft SUSY breaking terms. Usually it is assumed the
gravitino is heavier than the lightest neutralino $m_{3/2}> m_{\tz_1}$, in which case
the gravitino essentially decouples from phenomenology. However, if $m_{3/2}<m_{\tz_1}$, 
then the $\tz_1$ becomes unstable and can decay via modes such as $\tz_1\to \gamma\tG$.
The $\tz_1$ lifetime is expected to be very long-- of order $10^4-10^{12}$ sec-- 
so the neutralino still escapes detection at collider experiments, but is susceptible to
constraints from Big Bang nucleosynthesis (BBN) and CMB anisotropies~\cite{swimp}.
The relic density of gravitinos is expected to be simply 
$\Omega_{\tG}=\frac{m_{3/2}}{m_{\tz_1}}\Omega_{\tz_1}h^2$, since the gravitinos ``inherit''
the thermally produced neutralino relic number density. Thus, a scenario with a
$\tG$ superWIMP as LSP in SUGRA-type models can reduce the relic density by typically
factors of a few-- which is not enough in the case of Yukawa-unified models, where
relic density suppression factors of $10^2-10^5$ are needed. 

A better option occurs if we hypothesize an axino $\ta$ LSP. If indeed there is a
Peccei-Quinn solution to the strong $CP$ problem, then one expects the existence
of axions, typically with mass below the eV scale. While axions can themselves form 
cold dark matter, it is also easily possible that they contribute little to the
CDM relic density. However, in models with SUSY {\it and} axions, then the axion
is just one element of an axion superfield, the superpartner of the axion 
being a spin-${1\over 2}$ axino $\ta$.
The axino mass can be far different from the typical soft SUSY breaking scale, 
and the range $m_{\ta}\sim\;$eV--GeV is allowed. 

Axinos can be produced in the early universe both
thermally or non-thermally from NLSP decay. 
From the latter source, we expect roughly~\cite{ckkr}
\be
\Omega_{\ta}\sim \frac{m_{\ta}}{m_{\tz_1}}\Omega_{\tz_1}h^2 .
\ee
Thus, for $\Omega_{\tz_1}\sim 10^3$ and with $m_{\tz_1}\sim 50$ GeV as in Yukawa-unified models, 
an axino mass of $m_{\ta}\alt 5$ MeV is required

In this mass range, the axinos from $\tz_1$ decay are expected to give a hot/warm component to the
dark matter~\cite{jlm}. However, thermally produced axinos in this mass range could 
yield the required cold dark matter. Thus, if an unstable neutralino decay
$\tz_1\to \ta\gamma$ is to reconcile Yukawa-unified models with the relic density, 
then we would expect the dark matter to be predominantly cold axinos
produced thermally, but with a re-heat temperature $T_R<T_f$, where $T_f$ is the
temperature where axinos decouple from the thermal plasma in the early universe.
This scenario admits a dark matter abundance that can be in accord with WMAP measurements,
and would be primarily CDM, but with a  warm dark matter component arising 
non-thermally from $\tz_1$ decays. 
For a bino-like neutralino, as in Yukawa-unified models, the $\tz_1$ lifetime is 
given by~\cite{covi_prl}
\be
\tau\simeq 3.3\times 10^{-2}{\rm sec}\frac{1}{C_{aYY}^2}\left(\frac{f_a/N}{10^{11}{\rm GeV}}\right)^2
\left(\frac{50\ {\rm GeV}}{m_{\tz_1}}\right)^3 ,
\ee
where the model-dependent constant $C_{aYY}$ is of order 1, $f_a$ is the Peccei-Quinn breaking scale, and 
$N$ is a model dependent factor ($N=1 (6)$ for the KSVZ (DFSZ) axion model).
Thus, for reasonable choices of model parameters, we expect 
the neutralino lifetime to be of order $3\times 10^{-2}$ sec. This is short enough so that 
photon injection into the early universe from $\tz_1\to \ta\gamma$ decay 
occurs {\it before} nucleosynthesis, thus avoiding constraints from BBN. 

For illustration, we adopt a point A listed in Table \ref{tab:bm} of Yukawa-unified
benchmark models. The point has $m_{16}=9202.9$ GeV, $m_{10}=10966.1$ GeV, 
$M_D=3504.4$ GeV, $m_{1/2}=62.5$ GeV, $A_0=-19964.5$ GeV, $\tan\beta =49.1$ GeV
with $\mu >0$ and $m_t=171$ GeV. It has $m_{\tz_1}=55.6$ GeV and $\Omega_{\tz_1}h^2=423$
(IsaReD result).
Thus, $\tz_1\to \ta\gamma$ with $m_{\ta}\alt 1$ MeV would allow for a mixed warm/cold
axino dark matter solution to the problem of relic density in Yukawa-unified models.

\subsubsection{Dark matter solution via non-universal gaugino masses}

An alternative solution to reconciling the dark matter abundance with Yukawa-unified
models is to consider the possibility of non-universal gaugino masses. If we adopt any of the 
Yukawa unified models from the random scan and vary the $SU(2)$ ($SU(3)$) gaugino masses
$M_2$ ($M_3$), then the Yukawa coupling unification will be destroyed via the effect of
$\tst_i\tw_j$ ($\tg\tq$) loops. However, if $M_1$ is varied, Yukawa coupling unification
is preserved since contributions to fermion masses from loops containing $\tz_1$ are small.

By raising the GUT scale value of $M_1$ to values higher than $m_{1/2}$, the weak 
scale value of $M_1$ is also increased. If $M_1$ is increased enough, then 
$m_{\tz_1}$ (which is nearly equal to $M_1$ since $\tz_1$ is largely bino-like)
becomes close to $m_{\tw_1}$. When this happens, the $\tz_1$ becomes more wino-like,
with an increased annihilation cross section to $WW$ pairs if $m_{\tz_1}>M_W$~\cite{mwdm}. 
In our case, usually $m_{\tz_1}<M_W$. Then raising $M_1$ still lowers the relic density, 
but now via bino-wino co-annihilation (BWCA)~\cite{bwca}.  

In Fig.~\ref{fig:M1} we show the variation in $\Omega_{\tz_1}h^2$ versus
$M_1(M_{GUT})$ for benchmark point A in Table \ref{tab:bm}. The location of
$M_1$ for point A is marked by the arrow. The double dips at low $M_1$
are due to neutralino annihilation through the $Z$ and $h$ poles. 
Once $M_1(M_{GUT})$ is increased
to $\sim 195$ GeV, then we reach a relic density in accord with WMAP
measurements. Since $m_{\tw_1}\simeq m_{\tz_1}$, and $m_{\tw_1}\sim m_{\tz_1}$, 
the $\tz_2 -\tz_1$ mass gap is small, of order 10--20 GeV. We list the raised 
$M_1(M_{GUT})=195$ GeV point as point B in benchmark Table \ref{tab:bm}.

\FIGURE[t]{
\epsfig{file=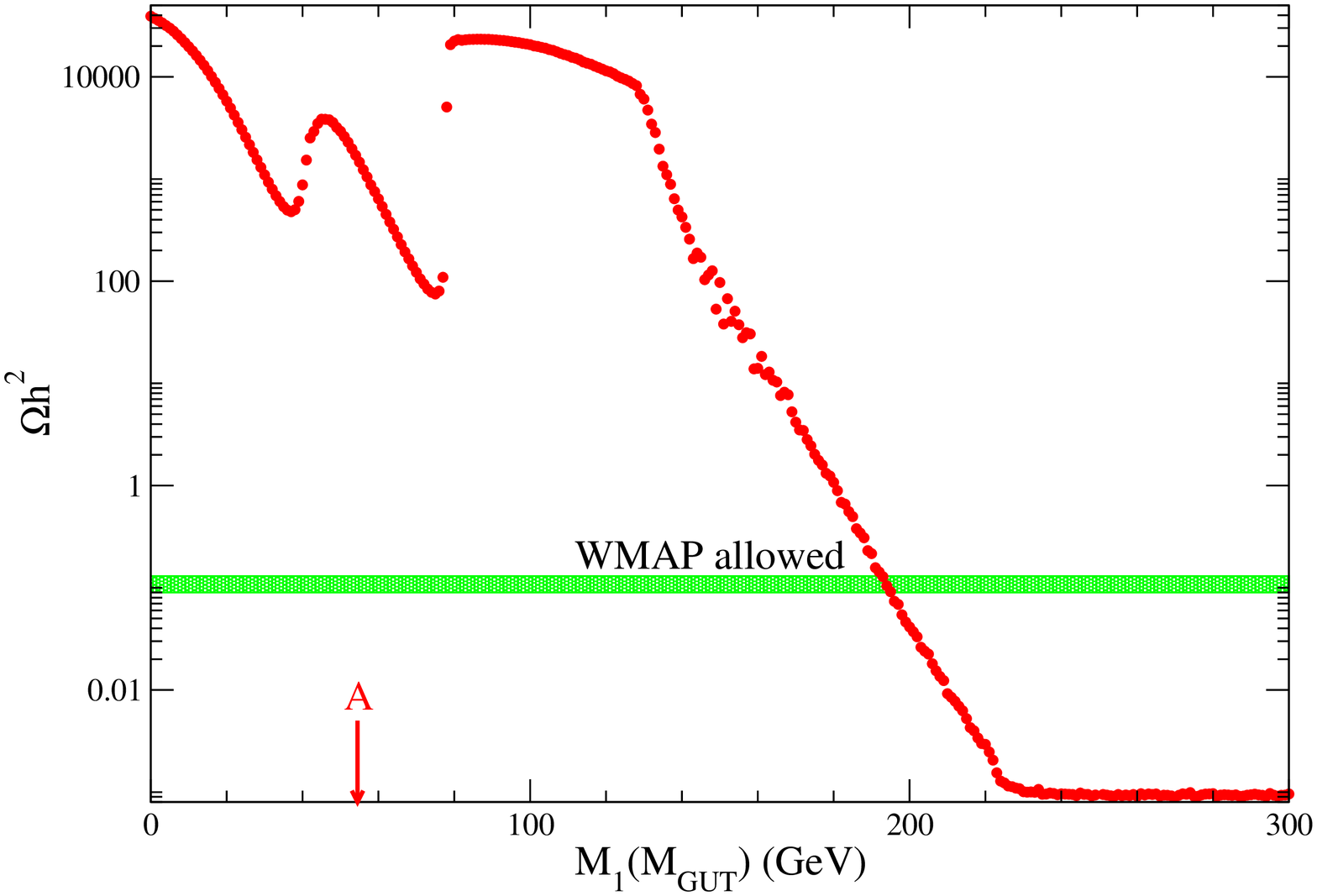,width=10cm} 
\caption{Plot of variation in $\Omega_{\tz_1}h^2$ versus non-universal 
GUT scale gaugino mass $M_1$ for benchmark point A in Table \ref{tab:bm}.}
\label{fig:M1}}

\subsubsection{Dark matter solution via generational non-universality}

Another possibility for reconciling the neutralino relic density with the 
measured value is to lower the first/second generation scalar masses 
$m_{16}(1,2)$, while keeping $m_{16}(3)$ fixed at $m_{16}$. 
The Bagger {\it et al.} inverted hierarchy solution depends only on
third generation scalar masses, while the effects of the first two generations decouple.
Ordinarily, solutions with $m_{16}(1,2)=m_{16}(3)$ are taken to enforce the super-GIM
mechanism for suppression of flavor changing neutral current (FCNC) processes. 
Limits from FCNCs mainly require near degeneracy between the first two generations, while limits on
third generation universality are much less severe~\cite{masiero}.
Lowering $m_{16}(1,2)$ works to lower the relic density because of the large $S$ term
in the scalar mass RGEs:
\be
S= m_{H_u}^2-m_{H_d}^2+Tr\left[{\bf m}_Q^2-{\bf m}_L^2-2{\bf m}_U^2+{\bf m}_D^2+{\bf m}_E^2\right] .
\ee
In models with universality, like mSUGRA,  
$S=0$ to one-loop at all energy scales; in models with non-universal Higgs scalars, 
like the HS model, this term can be large and have a major influence on scalar mass running.
The large $S$ term helps suppress right-squark masses. If $m_{16}(1,2)$ is taken light enough, 
then $m_{\tu_R}\simeq m_{\tc_R}\simeq m_{\tz_1}$, and neutralino-pair annihilation into quarks  
and neutralino-squark co-annihilation can act to reduce the relic density. 

In Fig.~\ref{fig:m16_12}, we show the variation in $\Omega_{\tz_1}h^2$ versus 
$m_{16}(1,2)$ where we take $m_{16}(3)= 5018.8$ GeV, $m_{1/2}=160$ GeV, 
$A_0=-10624.2$ GeV, $\tan\beta =47.8$ and $\mu >0$.
When $m_{16}(1,2)$ is lowered to $603.8$ GeV, then $m_{\tu_R}\simeq m_{\tc_R}=98.3$ GeV,
and we have neutralino annihilation via light $t$-channel squark exchange
and also neutralino-squark co-annihilation.\footnote{
A bug fix is needed in the Isajet 7.75 IsaReD subroutine in order to obtain the 
correct relic density; we thank A. Pukhov for pointing this out.}
IsaReD and Micromegas give $\Omega_{\tz_1}h^2\sim 0.1$ at this point, which we adopt as benchmark 
point C in Table \ref{tab:bm}.
The two light squarks are just at the limit of LEP2 exclusion. They may possibly
be excludable by Tevatron analyses, but the squark-neutralino mass gap is quite small, so the energy 
release from $\tu_R\to u\tz_1$ is low. So far, no such study has been made, 
and so the possibility cannot yet be definitively excluded.

\FIGURE[t]{
\epsfig{file=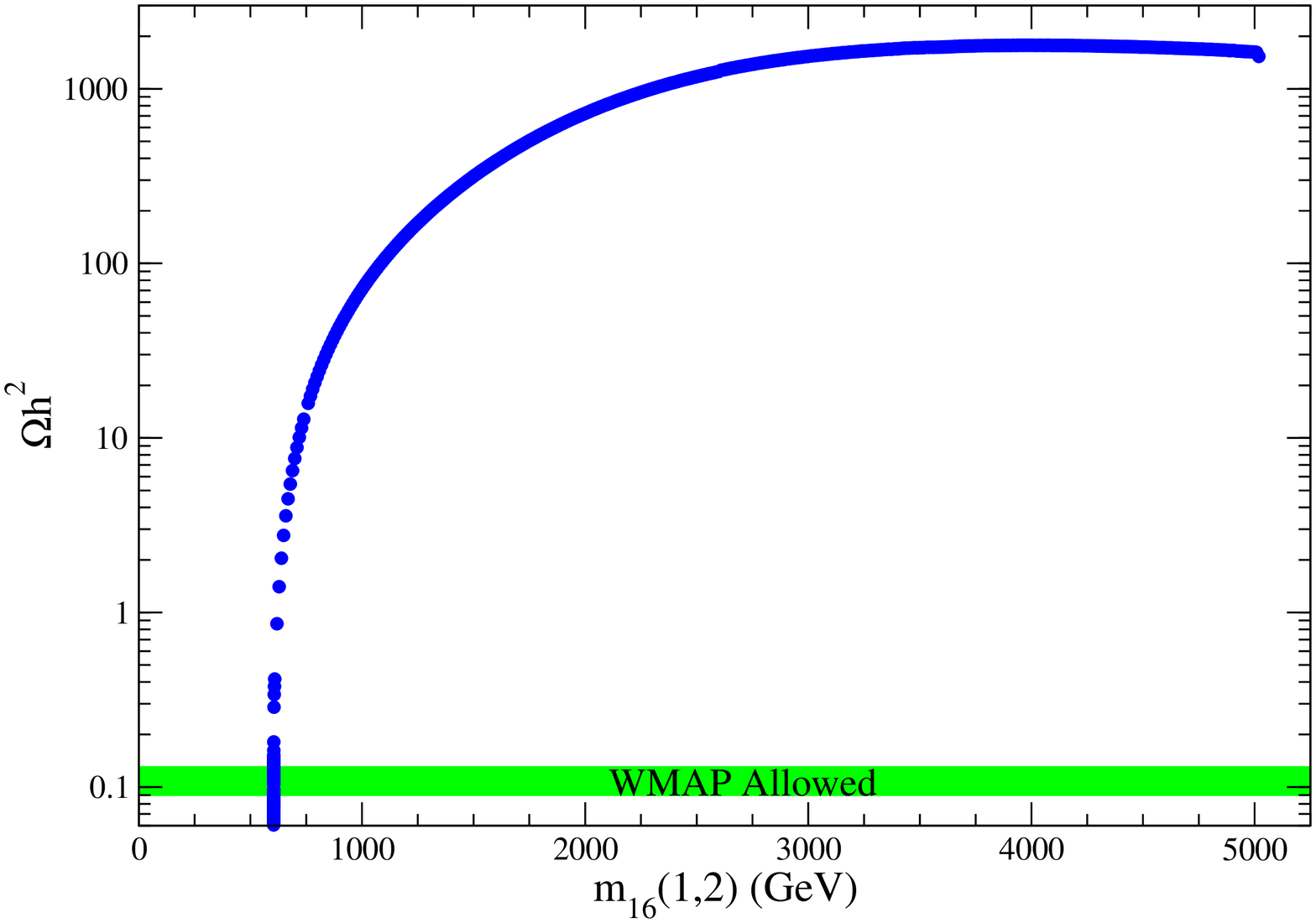,width=10cm} 
\caption{Plot of variation in $\Omega_{\tz_1}h^2$ versus non-universal 
GUT scale first/second generation scalar mass $m_{16}(1,2)$ 
for benchmark point C in Table \ref{tab:bm}.
}
\label{fig:m16_12}}

\clearpage
\section{Markov Chain Monte Carlo analysis}\label{sec:mcmc}

Next we adopt an improved scanning method based on the Markov Chain Monte Carlo 
(MCMC) technique to search more efficiently for parameter space regions of good Yukawa unification and 
WMAP-compatible DM relic density. A Markov Chain~\cite{mcmc} is a discrete-time, random process having the Markov 
property, which is defined such that given the present state, the future state only depends on the 
present state, but not on the past states. That is:  

\be
P(X^{t+1} = x|X^t = x^t,...,X^1=x^1)=P(X^{t+1}=x|X^t=x_t).
\label{eq:mkch}
\ee

An MCMC constructs a Markov Chain through sampling from a parameter space with the help of a specified 
algorithm.  In this study, we have applied the Metropolis-Hastings algorithm~\cite{metropolis}, which generates a 
candidate state $x^c$ from the present state $x^t$ using a proposal density $Q(x^t;x^c)$.  The candidate state is 
accepted to be the next state $x^{t+1}$ if the ratio

\be
p = \frac{P(x^c)Q(x^t;x^c)}{P(x^t)Q(x^c;x^t)},
\label{eq:pratio}
\ee
(where $P(x)$ is the probability calculated for the state $x$) is greater than a uniform random 
number $a = U(0,1)$.  If the candidate is not accepted, the present state $x^t$ is retained and a new 
candidate state is generated.  For the proposal density we use a Gaussian distribution that is centered 
at $x^t$ and has a width $\sigma$. This simplifies the $p$ ratio to $P(x^c)/P(x^t)$.  

Once taking off from a starting point, Markov chains are aimed to converge at a target distribution 
$P(x)$ around a point with the highest probability. The time needed for a Markov chain to converge depends 
on the width of the Gaussian distribution used as the proposal density. This width can be adjusted 
during the run to achieve a more efficient convergence.

In our search in the $SO(10)$ parameter space, we assume flat priors and we approximate the likelihood  
of a state to be $e^{-\chi^2(x)}$.  We define the $\chi^2$ for $R$ as
\be
\chi^2_R = \left(\frac{R(x)-R_{unification}}{\sigma_R}\right)^2
\label{eq:chi2R}
\ee
where $R_{unification}=1$ and $\sigma_R$ is the discrepancy we allow from absolute Yukawa unification, 
which, in this case we take to be $0.05$.  On the other hand, for $\Omega h^2$ we define 
\be
\chi^2_{\Omega h^2} = \left\{
      \begin{array}{lll}
      1, & {\hspace*{0.4cm}} & (0.094 \leq \Omega h^2 \leq 0.136)\\
      \left(\frac{\Omega h^2(x) - \Omega h^2_{mean}}{\sigma_{\Omega h^2}}\right)^2, & & 
	(\Omega h^2 < 0.094~{\rm or}~\Omega h^2 > 0.136)
      \end{array}
    \right. 
\label{eq:chi2omg}
\ee
where $\Omega h^2_{mean}=0.115$ 
is the mean value of the range $0.094<\Omega h^2<0.136$ proposed in~\cite{hamann}, and 
$\sigma_{\Omega h^2}=0.021$.   
This way, the MCMC primarily searches for regions of Yukawa-unifications, and within these 
regions for solutions with a good relic density. 

For each search, we select a set of 
$\sim 10$ starting points in order to ensure a more thorough investigation of the parameter space.  Then 
we run the MCMC, aiming to maximize the likelihood of either $R$ alone, or $R$ and $\Omega h^2$ 
simultaneously.  For the case of simultaneous maximization, we compute the $p$ ratios for $R$ and 
$\Omega h^2$ individually, requiring both $p_R > a$ and $p_{\Omega h^2} > a$ separately.  We do not 
strictly seek convergence to an absolute maximal likelihood, but we rather use the MCMC as a tool to 
reach compatible regions and to investigate the amount of their extension in the $SO(10)$ parameter 
space.

\subsection{HS model: neutralino annihilation via $h$ resonance}
\label{sec:h}

We begin our MCMC scans by selecting 10 starting points ``pseudorandomly" --that is, selecting them 
from different $m_{16}$ regions to cover a wider range of the parameter space-- and imposing some loose 
limits (defined by our previous works and random scans) on the rest of their parameters to achieve a 
more efficient convergence. Our initial scan  is directed to look for points only with 
$R$ as close to 1.0 as possible by maximizing solely the likelihood of $R$. Based on the results of the 
first MCMC scan, we then pick a new set of 10 starting points with low $R$ and also low 
$\Omega_{\tz_1}h^2$, and direct the second scan to look for points with both $R=1.0$ and 
$\Omega_{\tz_1}< 0.136$ by maximizing the likelihoods of $R$ and $\Omega h^2$ simultaneously. For MCMC 
scans, the code is interfaced to the micrOMEGAs~\cite{micromegas} package to evaluate the relic density 
and low-energy constraints. 

Figure~\ref{fig:pm16m10HS} shows the Yukawa-unified region found by the MCMC results 
as a projection in the plane of $m_{16}$ versus $m_{10}$. The light-blue dots are points which
have $R<1.1$, while dark blue dots have $R<1.05$. In addition, we show in orange (red) 
the points which satisfy $R<1.1$ ($1.05$) {\it and} $\Omega_{\tz_1}h^2 <0.136$. 
The points with low $R$ are narrowly correlated along the line 
$m_{10}\simeq 1.2 m_{16}$. While the low $R$ points range over $m_{16}$ values
from 3 to over 12 TeV (in agreement with the results from the random scans) the MCMC 
has also identified a range of points with {\it both} $R\simeq 1$ and $\Omega_{\tz_1}h^2<0.136$, 
but only for $m_{16}$ values of about 3--4 TeV!

Fig.~\ref{fig:pm16a0Dm16HS} shows the MCMC scan results in the $m_{16}$~vs.~$A_0/m_{16}$
plane. Again, we see that points with low $R$ populate the region with $A_0\sim (2\textrm{--}2.1) m_{16}$
over a wide range of $m_{16}$ values. The plot includes the $\Omega_{\tz_1}h^2<0.136$
points around $m_{16}\sim 3$--$4$ TeV.

In Fig.~\ref{fig:pm16mhfHS}, we show MCMC results in the $m_{16}$~vs.~$m_{1/2}$ plane.
Here, we see the very lowest $R$ points select out the lowest possible $m_{1/2}$ values
allowed for a given value of $m_{16}$, and that the minimum $m_{1/2}$ value allowed
steadily decreases with increasing $m_{16}$-- the boundary being determined by the LEP2 
limit on chargino masses. The points with a ``good" relic density are clustered 
around $m_{1/2}\sim 100$ GeV.

We also show in Fig.~\ref{fig:pm16mHHS} the individual GUT-scale values of Higgs soft 
terms $m_{H_u}$ (lower branch) and $m_{H_d}$ (upper branch). This plot displays the
required Higgs splitting and confirms that $m_{H_d}>m_{H_u}$.

In Fig.~\ref{fig:pmh0Mm2ne1ma0Mm2ne1HS}, we show points with low $R$ in the
$m_h-2m_{\tz_1}$~vs.~$m_A-2m_{\tz_1}$ plane. In these solutions, 
$m_A$ is usually far greater than $2m_{\tz_1}$, indicating the 
neutralino annihilation through the $A$-resonance is not the cause of the
reduced relic density orange and red points. 
However, the low $\Omega_{\tz_1}h^2$ points all {\it do}
lie along the $m_h\simeq 2m_{\tz_1}$ line, indicating that $h$-resonance 
annihilation is the mechanism at work to reduce the relic density in the 
early universe. In Fig. \ref{fig:pRomgHS}, we show $R$~vs.~$\Omega_{\tz_1}h^2$ for the MCMC
scan. In this frame, we see that the points with high relic density extend down to $R=1$, 
while the low relic density points reach below $R=1.05$, but can reach no lower than $R=1.03$.

\FIGURE[tbp]{
\epsfig{file=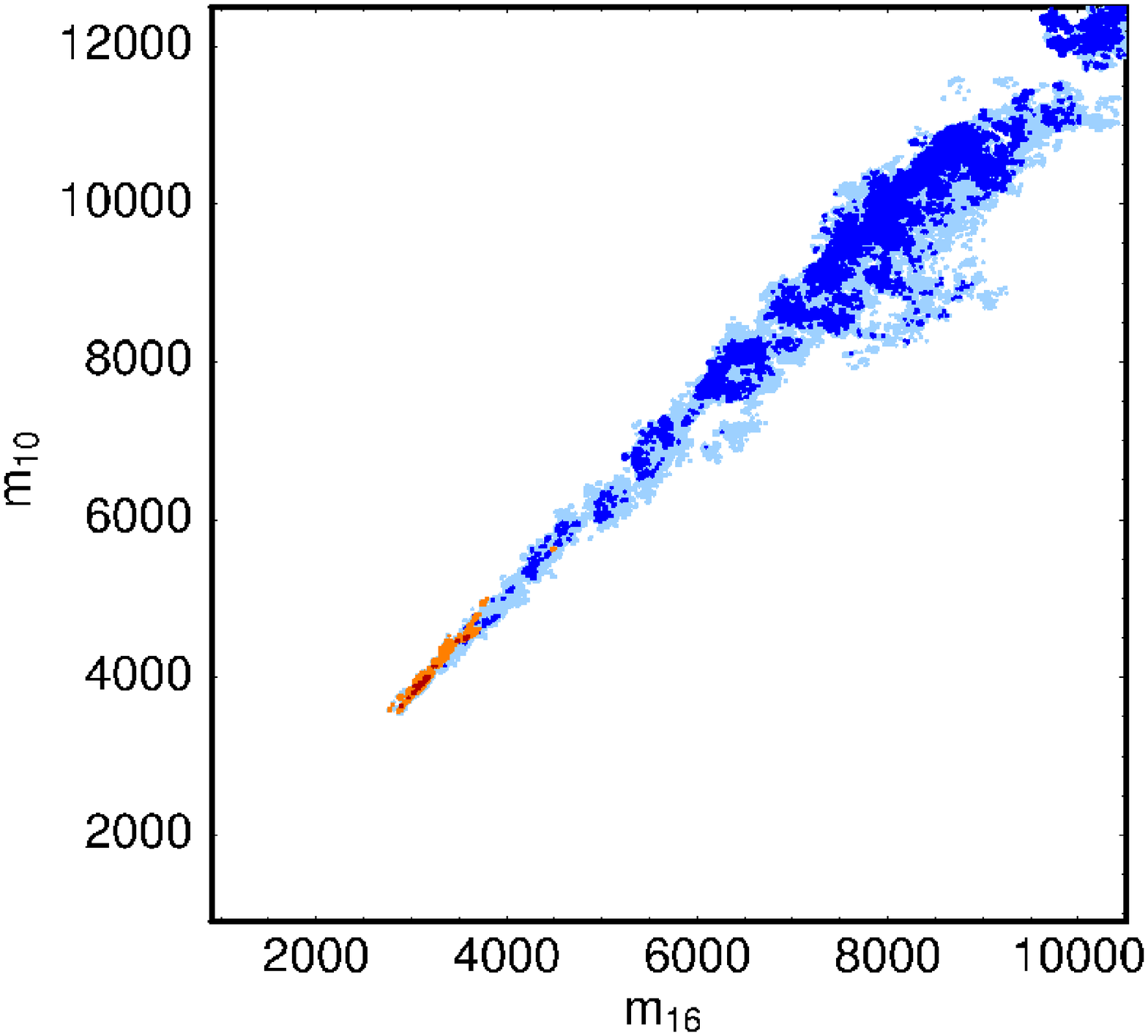,width=9cm} 
\caption{Plot of MCMC results in the $m_{16}$ vs. $m_{10}$
plane; the light-blue (dark-blue) points have $R<1.1\ (1.05)$, 
while for the orange (red) points $R<1.1\ (1.05)$ and $\Omega_{\tz_1}h^2<0.136$.}
\label{fig:pm16m10HS}}

\FIGURE[tbp]{
\epsfig{file=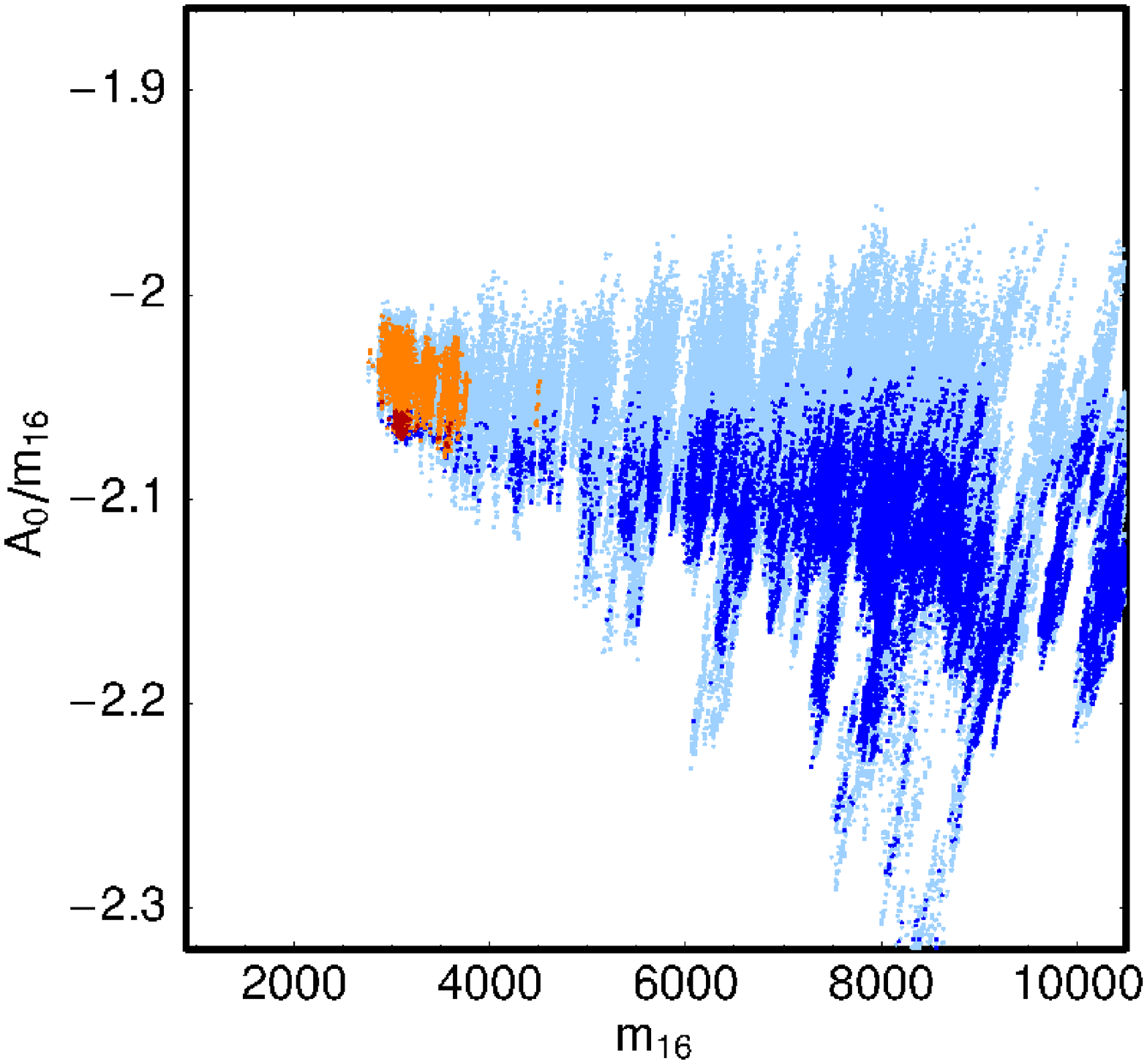,width=8.6cm} 
\caption{Plot of MCMC results in the $m_{16}$ vs. $A_0/m_{16}$ 
plane; the light-blue (dark-blue) points have $R<1.1\ (1.05)$, 
while for the orange (red) points $R<1.1\ (1.05)$ and $\Omega_{\tz_1}h^2<0.136$.}
\label{fig:pm16a0Dm16HS}}

\FIGURE[tbp]{
\epsfig{file=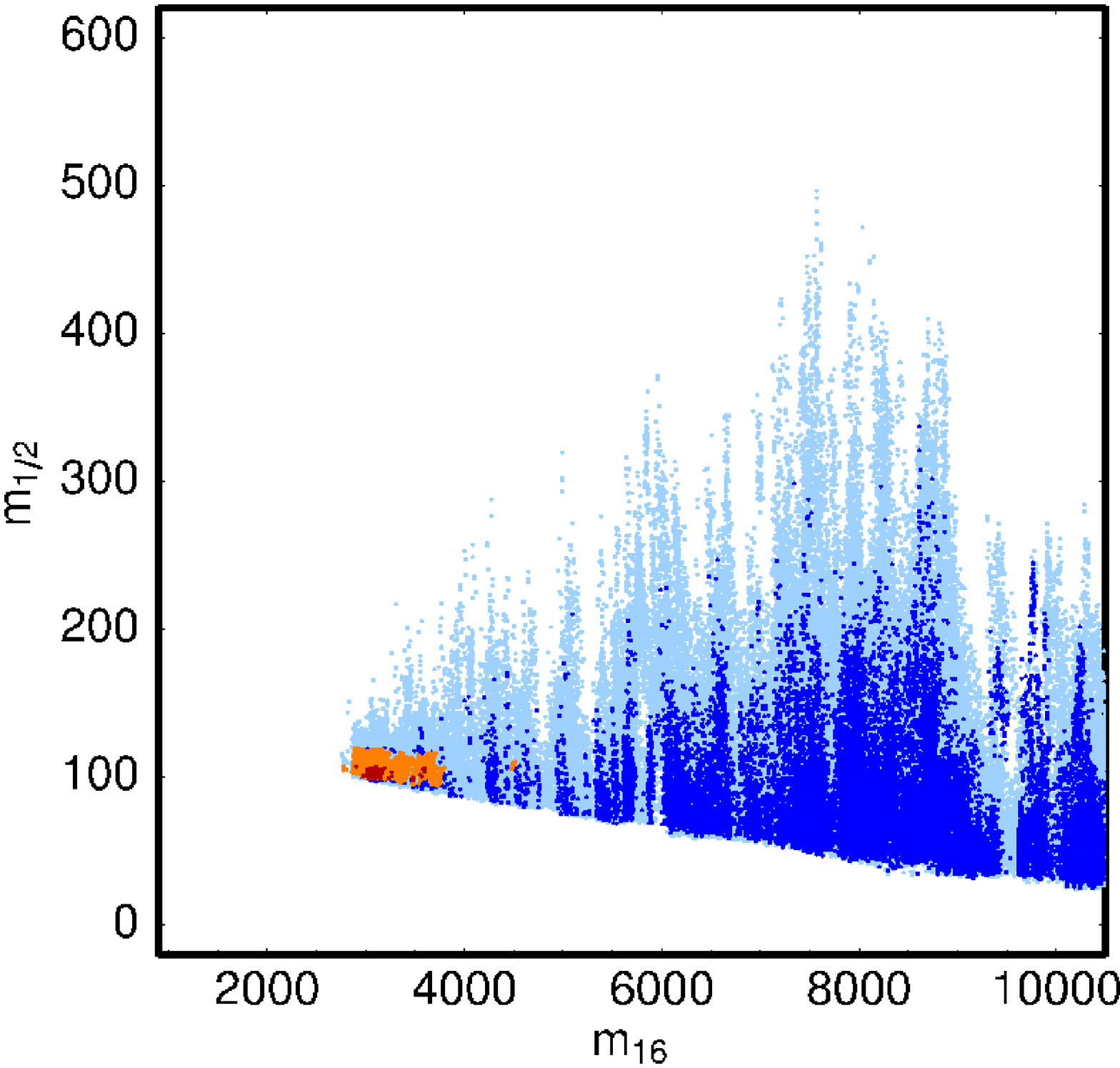,width=9cm} 
\caption{Plot of MCMC results in the $m_{16}$ vs. $m_{1/2}$
plane; the light-blue (dark-blue) points have $R<1.1\ (1.05)$, 
while for the orange (red) points $R<1.1\ (1.05)$ and $\Omega_{\tz_1}h^2<0.136$.}
\label{fig:pm16mhfHS}}

\FIGURE[tbp]{
\epsfig{file=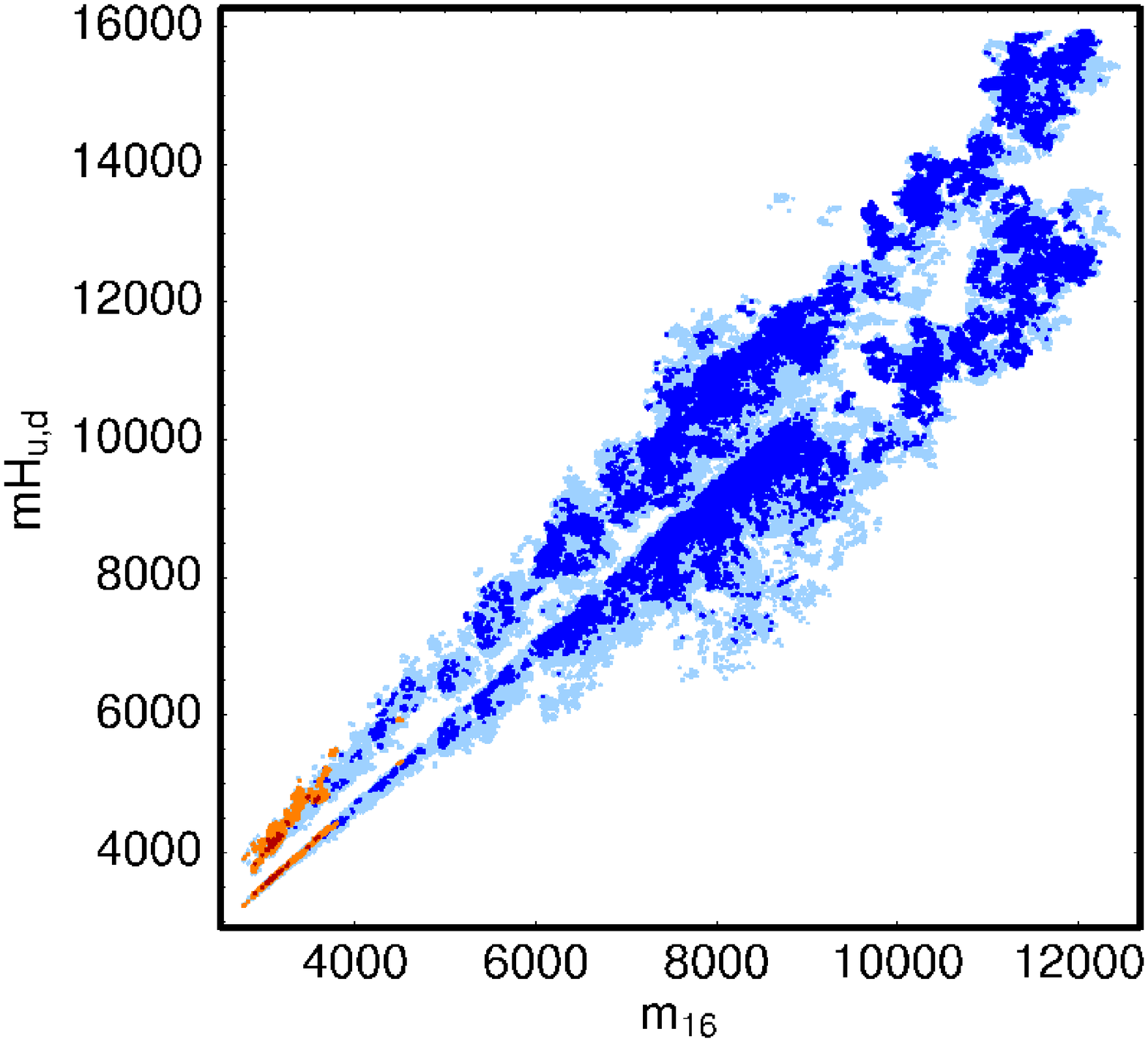,width=9.4cm} 
\caption{Plot of MCMC results in the $m_{16}$ vs. $m_{H_{d,u}}$
plane; the light-blue (dark-blue) points have $R<1.1\ (1.05)$, 
while for the orange (red) points $R<1.1\ (1.05)$ and $\Omega_{\tz_1}h^2<0.136$.}
\label{fig:pm16mHHS}}

\FIGURE[tbp]{
\epsfig{file=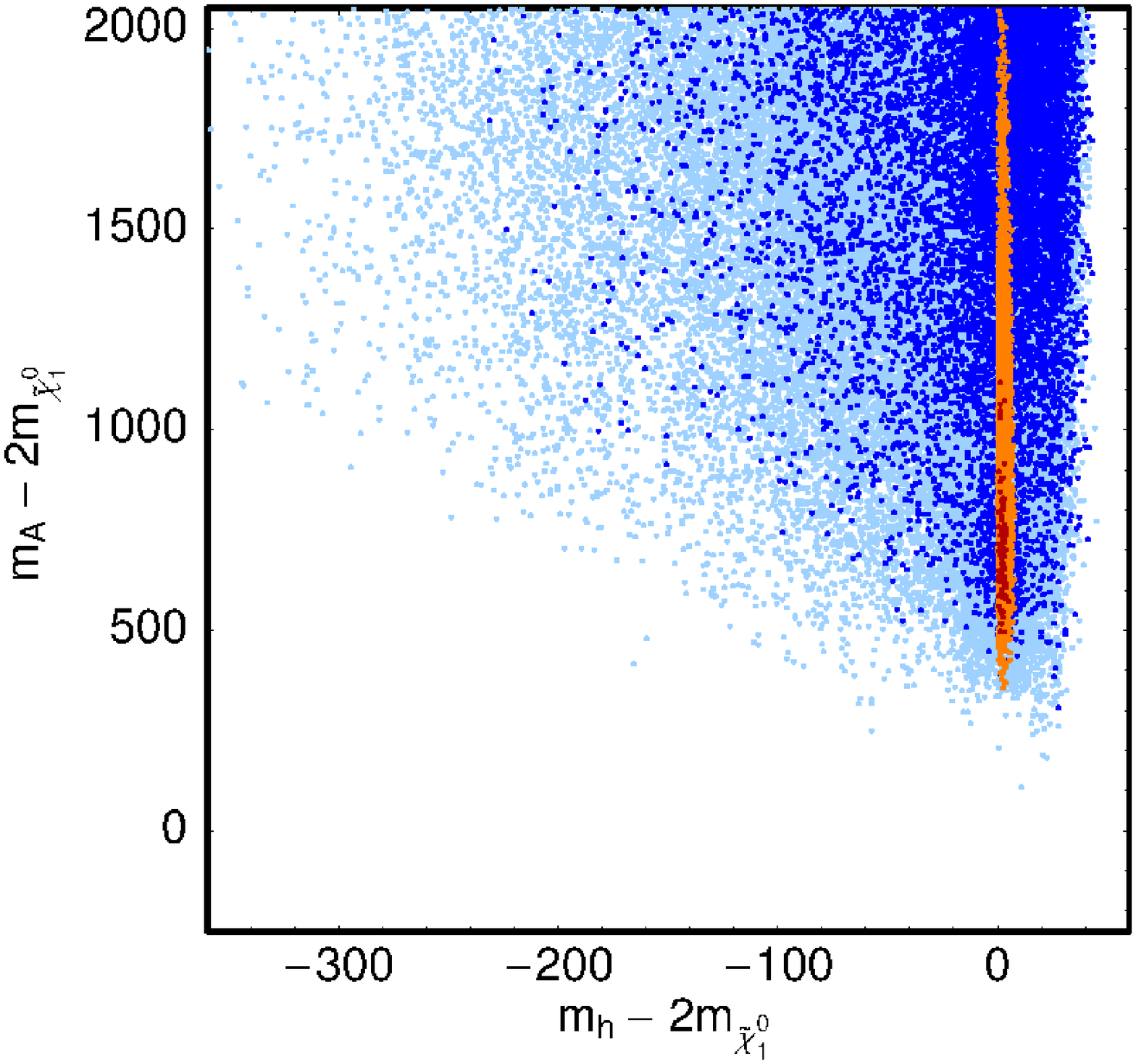,width=9cm} 
\caption{Plot of MCMC results in the $m_h-2m_{\tz_1}$ vs. $m_A-2m_{\tz_1}$
plane; the light-blue (dark-blue) points have $R<1.1\ (1.05)$, 
while for the orange (red) points $R<1.1\ (1.05)$ and $\Omega_{\tz_1}h^2<0.136$.}
\label{fig:pmh0Mm2ne1ma0Mm2ne1HS}}

\FIGURE[tbp]{
\epsfig{file=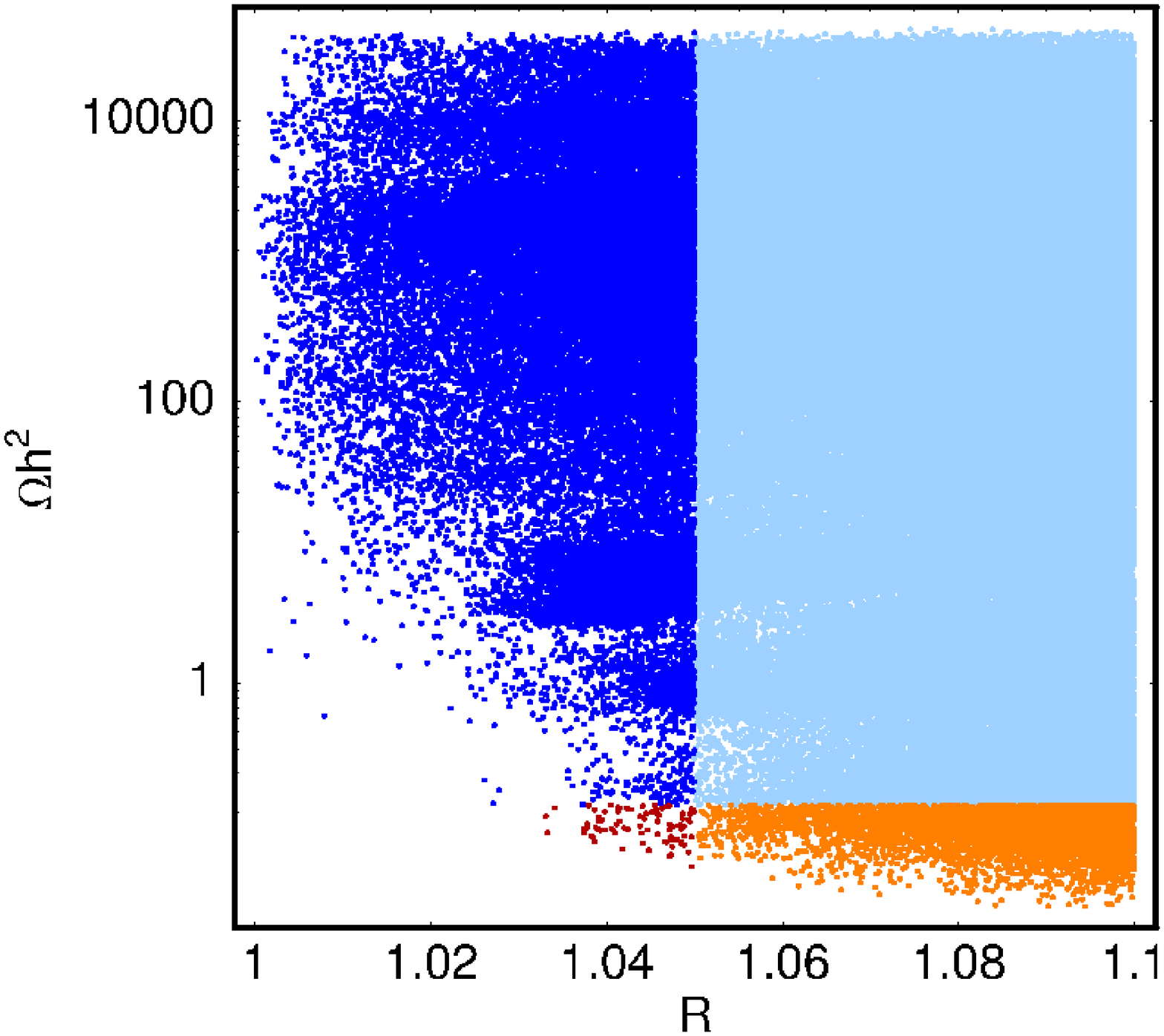,width=9cm} 
\caption{Plot of MCMC results in the $R$~vs.~$\Omega_{\tz_1}h^2$
plane; the light-blue (dark-blue) points have $R<1.1\ (1.05)$, 
while for the orange (red) points $R<1.1\ (1.05)$ and $\Omega_{\tz_1}h^2<0.136$.}
\label{fig:pRomgHS}}

We are now in a position to understand this new class of Yukawa-unified, DM-allowed solutions:
the search for low $R$ pushes $m_{16}$ to very high, multi-TeV values. Meanwhile, in order
for $h$-resonance annihilation to reduce the relic density to the WMAP-allowed range, 
$m_{16}$ can't be too large. The region around $m_{16}\sim 3$--$4$~TeV offers a 
compromise between these two tendencies: for $m_{16}$ not too large, the 
dip in relic density due to the $h$-resonance annihilation is sufficient to bring
the relic density into the desired range. But since $m_{16}$ can't be too large, 
the Yukawa unification is limited to a couple of percent at best.
This new class of solutions was difficult to reach using a random scan, since the
$h$-resonance is so narrow. The necessary value of $m_{\tz_1}$ has to be just right-- 
with $2m_{\tz_1}$ slightly below $m_h$-- so that the thermal averaging of 
neutralino energies convolutes with the resonant cross section with enough strength
to give substantial neutralino annihilation in the early universe.

The $SO(10)$ model parameters leading to low $R$ and good relic density occur only
over a very narrow range of $m_{1/2}\sim 100$ GeV and $m_{16}\sim 3$~TeV. This means the
Yukawa-unified $h$-resonance annihilation points have very specific mass spectra predictions.
We show in Fig.~\ref{fig:pmgmt1HS} the $m_{\tg}$~vs.~$m_{\tst_1}$ plane for MCMC points
in the HS model. Here, we see that the points with $\Omega_{\tz_1}h^2<0.136$ 
all have $m_{\tg}\sim 350$--$450$~GeV, while $m_{\tst_1}\sim 350$--$750$~GeV. 
The large $\mu$ parameter combines with gaugino mass unification to predict that 
$m_{\tw_1}\simeq m_{\tz_2}\sim 100$--$150$ GeV, while $m_{\tz_1}\sim 50$--$75$ GeV. 

Given this very tightly correlated mass spectrum, gluino cascade decay events at the
LHC will lead to $\tz_2$ production, followed mainly by $\tz_2\to \tz_1 b\bar{b}$ 
decay and also by $\tz_2\to \tz_1\ell\bar{\ell}$ decay.
The OS/SF isolated dilepton mass spectrum will be bounded kinematically
by $m_{\tz_2}-m_{\tz_1}$. The mass difference provides an edge in the dilepton mass spectrum which
is characteristic of these decays, and which is easily measureable. Furthermore, it should be
correlated with $m_h$, since all these masses are related by resonance annihilation and theory.
The value of $m_h$ should be directly measureable at LHC after several years 
of data taking via the bump in the
$h\to\gamma\gamma$ mass spectrum. Thus, we plot in Fig.~\ref{fig:pmne2Mmne1mh0HS} our
Yukawa-unified MCMC points in the $m_{\tz_2}-m_{\tz_1}$~vs.~$m_h$ plane. For this scenario to be
bourne out, we would predict $m_{\tz_2}-m_{\tz_1}\sim 52$--$65$~GeV, with a roughly linear
correlation with $m_h$. 

We adopt point D in Table \ref{tab:bm} as being representative of the light Higgs
$h$-resonance annihilation compromise solutions. 
The relic density computed with micrOMEGAs ($\Omega_{\tz_1}h^2=0.06$) 
is below the preferred range, while IsaReD gives $\Omega_{\tz_1}h^2=0.1$.
Yukawa couplings are unified at the 9\% level. We note here that
we could have adopted a solution with even better Yukawa coupling unification at the
4--5\% level. These solutions tend to give light Higgs mass $m_h\alt 110$ GeV 
(as can be seen by the red dots in Fig. \ref{fig:pmne2Mmne1mh0HS}) which are 
more likely to be excluded by LEP2 Higgs search results.

\FIGURE[t]{
\epsfig{file=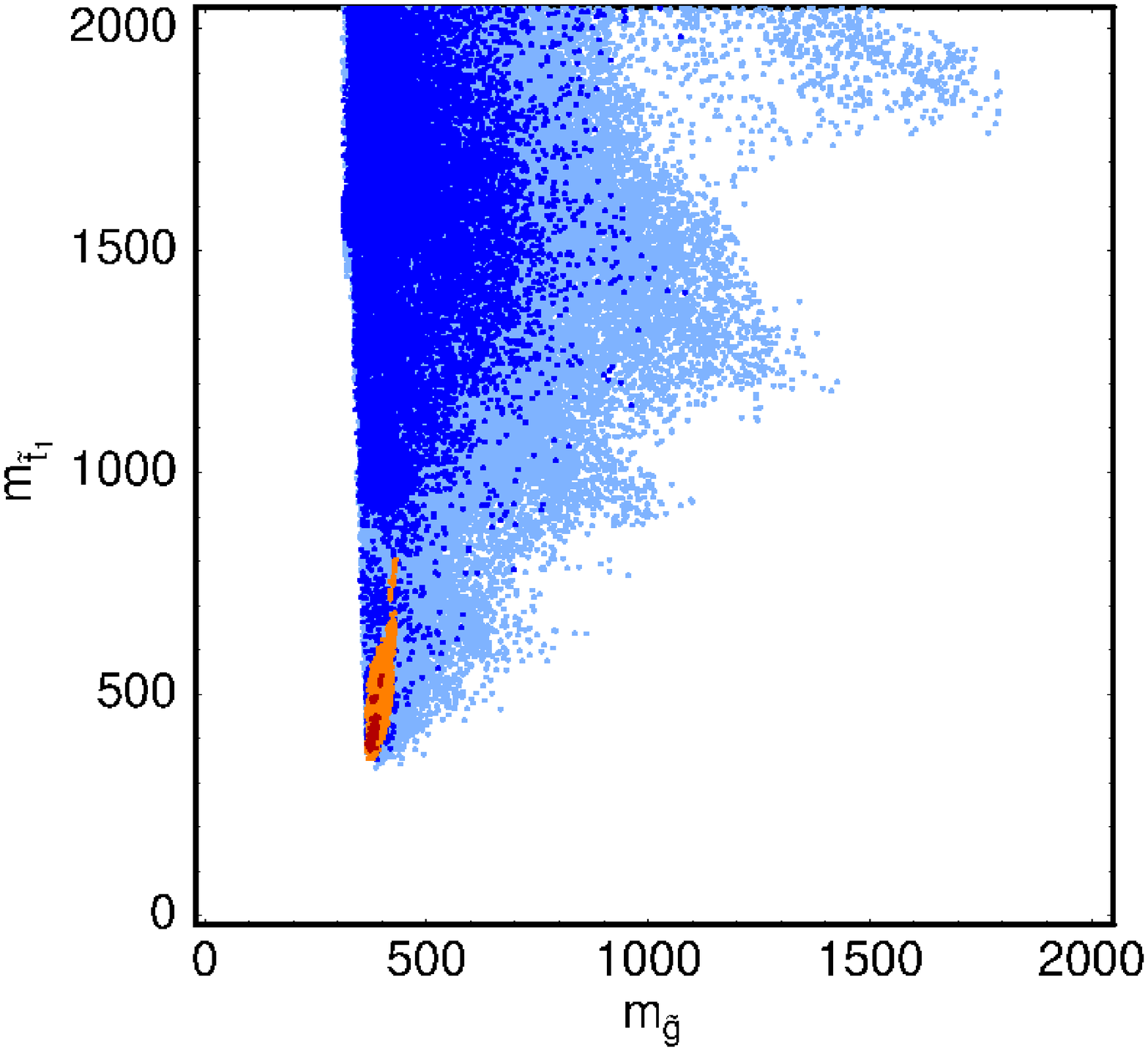,width=9cm} 
\caption{Plot of MCMC results in the $m_{\tg}$ vs. $m_{\tst_1}$
plane; the light-blue (dark-blue) points have $R<1.1\ (1.05)$, 
while for the orange (red) points $R<1.1\ (1.05)$ and $\Omega_{\tz_1}h^2<0.136$.}
\label{fig:pmgmt1HS}}

\FIGURE[tbp]{
\epsfig{file=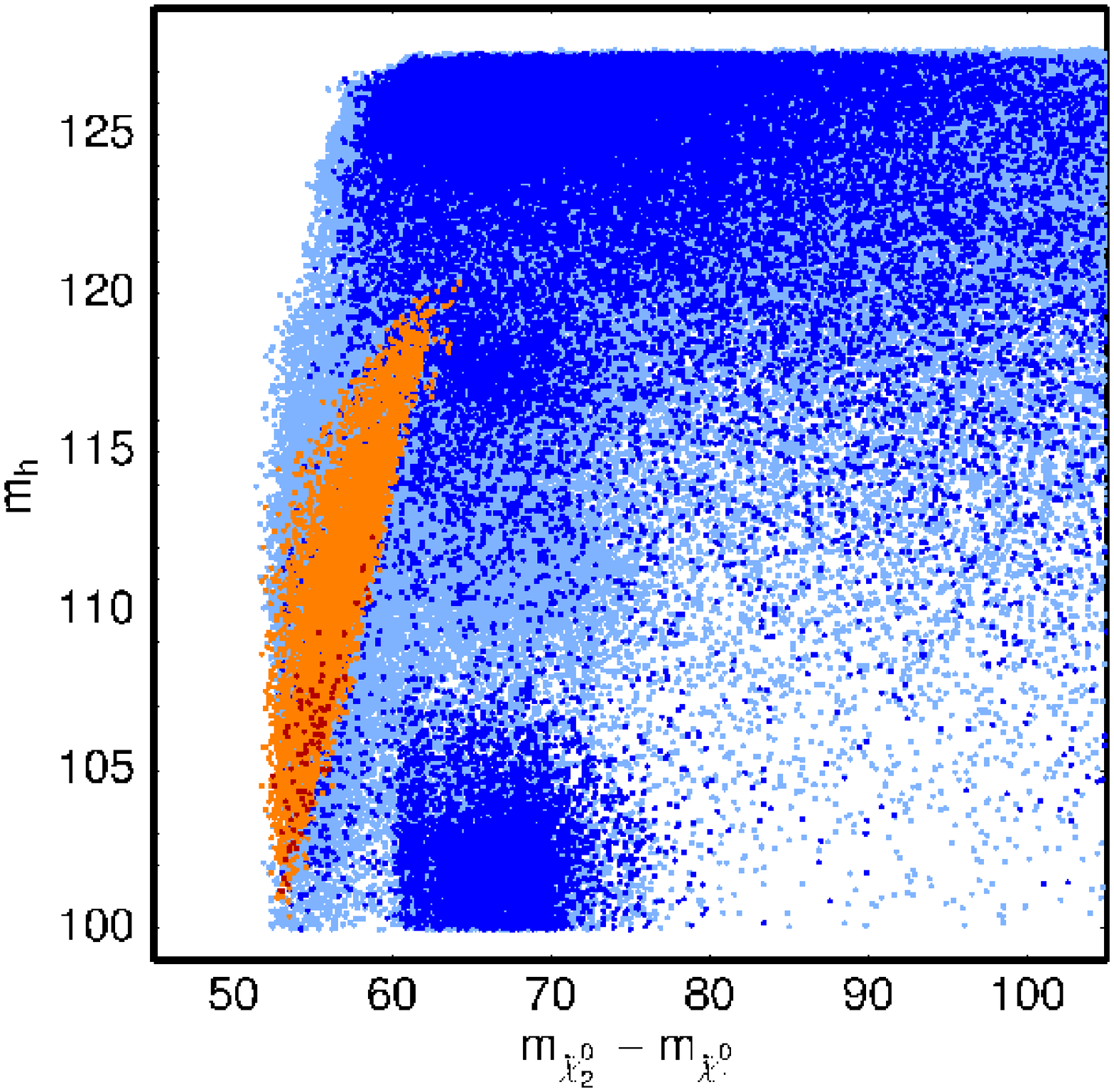,width=8.6cm} 
\caption{Plot of MCMC results in the $m_{\tz_2}-m_{\tz_1}$ vs. $m_h$
plane; the light-blue (dark-blue) points have $R<1.1\ (1.05)$, 
while for the orange (red) points $R<1.1\ (1.05)$ and $\Omega_{\tz_1}h^2<0.136$.}
\label{fig:pmne2Mmne1mh0HS}}

\clearpage

\subsection{Solutions using weak scale Higgs boundary conditions}
In the analysis put forth by BDR~\cite{bdr}, Yukawa-unified solutions are
found with low values of both $\mu$ and $m_A$ in the 100--200 GeV range, while $m_{16}$ and $m_{10}$
are typically at the 2--3 TeV scale. We have seen from our results so far that
$\mu$ and $m_A$ are typically in the TeV regime. Some low 
$\mu$ solutions were generated using Isajet in Table 2 of Ref.~\cite{abbbft}, but these had
$R\sim 1.25$. 

We find here that we can generate small $\mu$ and small $m_A$ solutions using Isajet
by using the pre-programmed non-universal Higgs model (NUHM)\footnote{This is  model line 8
of the Isajet non-universal supergravity models (NUSUG)}. The approach is to start with a
set of GSH soft term boundary conditions, and evolve the soft SUSY breaking Higgs masses
$m_{H_u}^2$ and $m_{H_d}^2$ down to the weak scale $M_{SUSY}$. At $Q=M_{SUSY}$, 
re-calculate what $m_{H_u}^2$ and $m_{H_d}^2$ {\it should have been} in order to get the input values
of $m_A$ and $\mu$, using the two electroweak symmetry breaking minimization conditions
(in practise, we use 1-loop relations):
\bea
B&=&\frac{(m_{H_u}^2+m_{H_d}^2+2\mu^2 )\sin 2\beta}{2\mu}\quad {\rm and}\\
\mu^2 &=&\frac{m_{H_d}^2-m_{H_u}^2\tan^2\beta}{(\tan^2\beta -1)}
-\frac{M_Z^2}{2} .
\label{eq:ewsb}
\eea

Then run back up to the GUT scale using these new WSH boundary conditions. At each iteration, the 
weak scale values of $m_{H_u}^2$ and $m_{H_d}^2$ have to be re-computed so as to 
maintain the input value of $\mu$ and $m_A$; in this case, the GUT scale values of $m_{H_u}^2$ and
$m_{H_d}^2$ are outputs, instead of inputs. For this class of solutions, 
both GSH {\it and} WSH boundary conditions must be used in Isajet. 
The GSH boundary conditions are needed just to get an acceptable EWSB on the first
iteration so that a spectrum can be computed, and then modified to yield the input
values of $m_A$ and $\mu$. 
Using default universal GSH soft terms will usually fail to give appropriate
EWSB on any iteration where Yukawa couplings are unified.

We implement an MCMC scan over the modified parameter space
\be
m_{16},\ m_{1/2},\ A_0,\ \tan\beta,\ m_A,\ \mu 
\ee
(effectively trading the GUT scale inputs $m_{H_u}^2$ and $m_{H_d}^2$ 
(or alternatively $m_{10}$ and $M_D^2$) for weak scale inputs $m_A$ and $\mu$). 
We begin with $10$ starting points selected pseudorandomly from different regions of the above 
parameter space, and implement two MCMC scans on them, 
one searching for points with lowest $R$ values 
by maximizing the likelihood of $R$ and the other for solutions with $R=1$ {\it and} 
$\Omega_{\tz_1}h^2<0.136$ by maximizing likelihoods of $R$ and $\Omega h^2$ simultaneously.

Our first results are shown in Fig.~\ref{fig:pm16a0Dm16BDR} for the
$m_{16}$~vs.~$A_0/m_{16}$ plane, where we plot
points with $R<1.1$ (1.05) using dark blue (light blue) dots, and solutions
with $\Omega_{\tz_1}h^2<0.136$ for $R<1.1$ (1.05) using orange (red) dots.
While we again get good Yukawa-unified solutions over a wide range of multi-TeV
values of $m_{16}$, this time we pick up {\it additional} dark matter allowed
solutions for $m_{16}:3$--$6$ TeV. The solutions again respect the Bagger {\it et al.}
boundary condition $A_0\simeq -2 m_{16}$.

\FIGURE[p]{
\epsfig{file=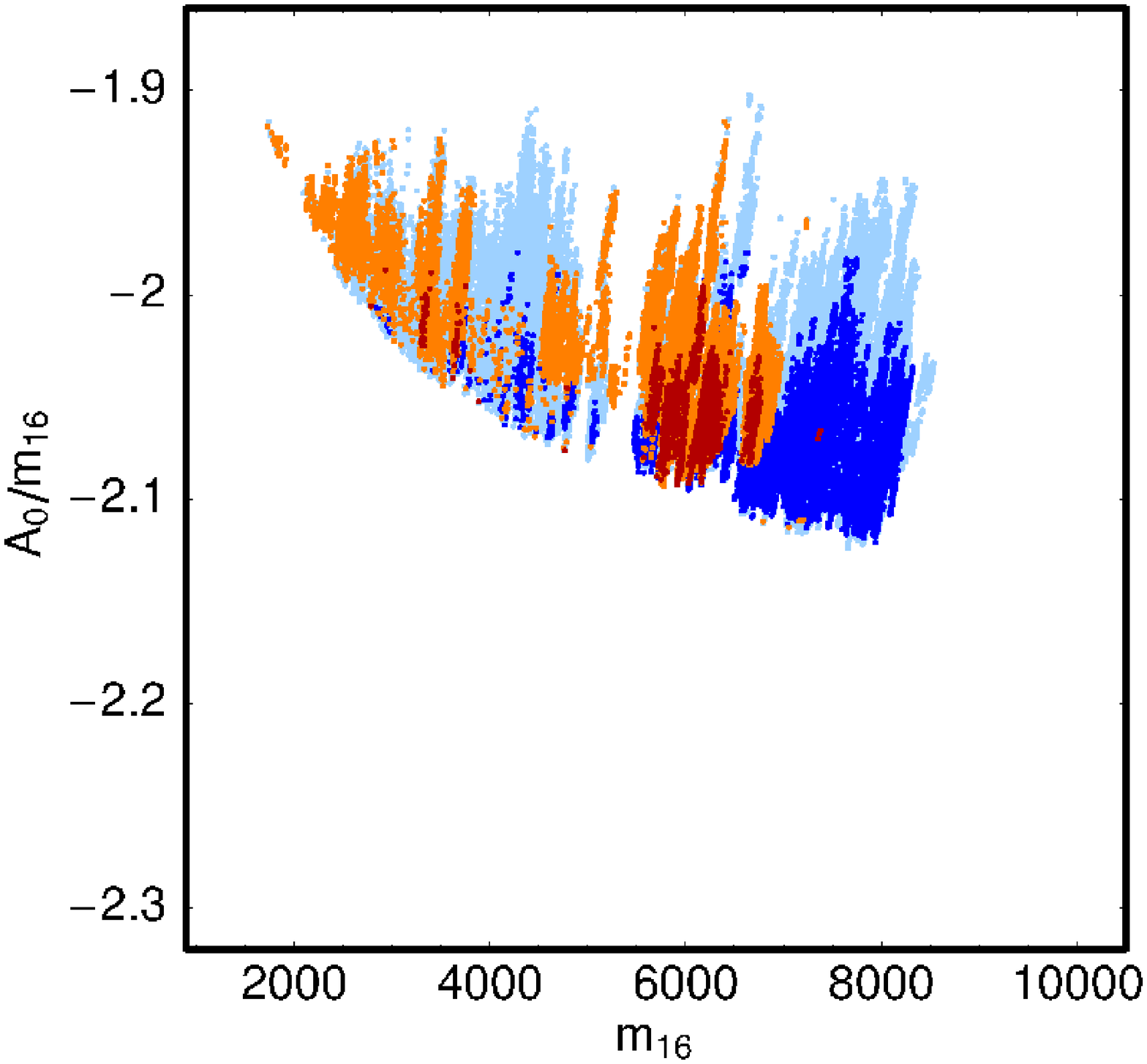,width=9cm} 
\caption{Plot of MCMC results using WSH boundary conditions 
in the $m_{16}$ vs. $A_0/m_{16}$
plane; the light-blue (dark-blue) points have $R<1.1\ (1.05)$, 
while for the orange (red) points $R<1.1\ (1.05)$ and $\Omega_{\tz_1}h^2<0.136$.}
\label{fig:pm16a0Dm16BDR}}

In Fig.~\ref{fig:pm16mhfBDR}, we show the WSH solutions in the $m_{16}$~vs.~$m_{1/2}$
plane. The minimum in allowed $m_{1/2}$ values again decreases with increasing
$m_{16}$. We see that for the WSH class of solutions, much larger values of $m_{1/2}$
ranging up to $300-500$ GeV are DM-allowed.

\FIGURE[p]{
\epsfig{file=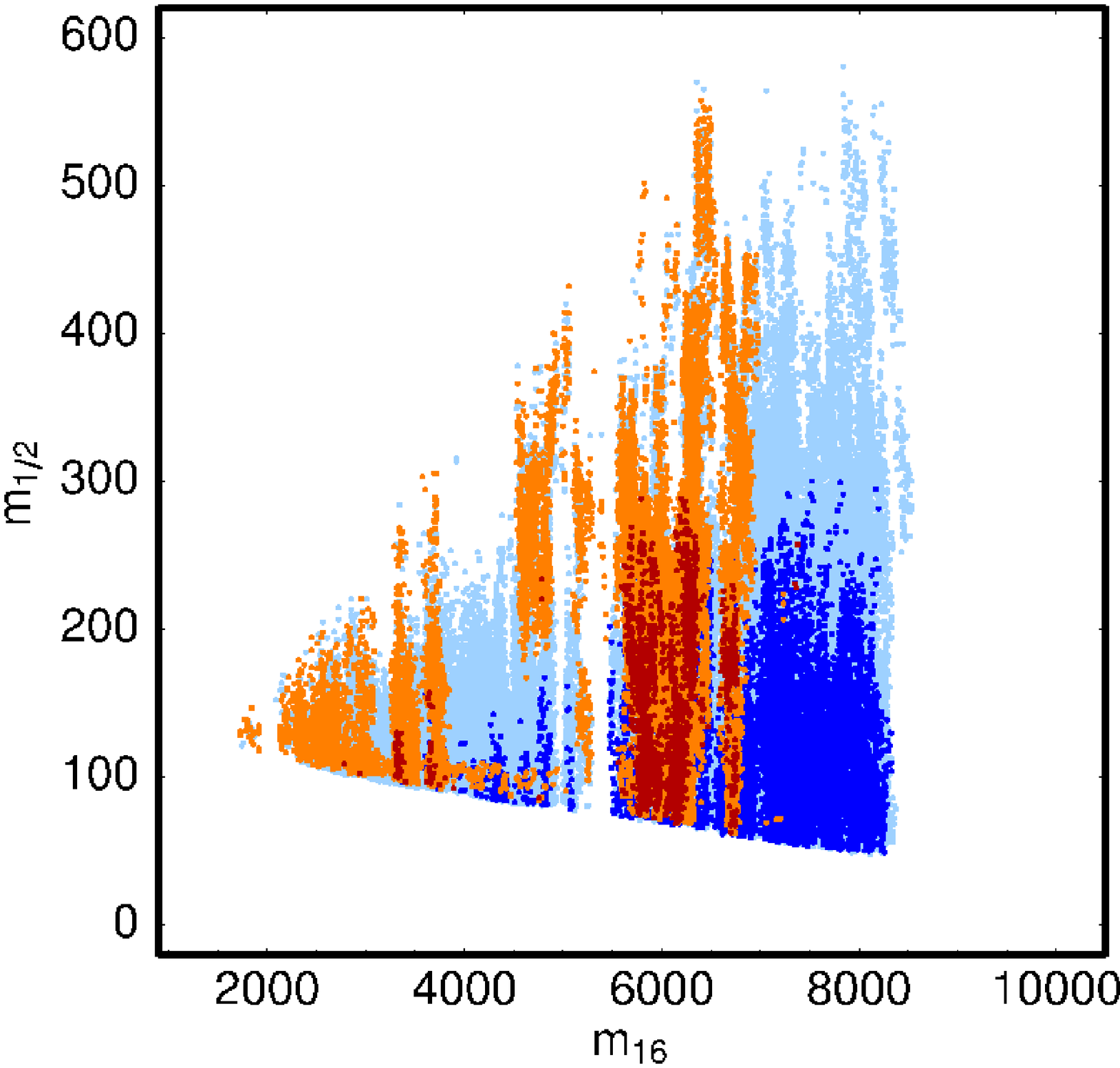,width=9cm} 
\caption{Plot of MCMC results using WSH boundary conditions 
in the $m_{16}$ vs. $m_{1/2}$ plane; the light-blue (dark-blue) points have $R<1.1\ (1.05)$, 
while for the orange (red) points $R<1.1\ (1.05)$ and $\Omega_{\tz_1}h^2<0.136$.}
\label{fig:pm16mhfBDR}}

In Fig.~\ref{fig:pma0muBDR}, we plot the WSH solutions in the input parameter
$m_A$~vs.~$\mu$ plane. In this case, we see that the bulk of the DM-allowed
solutions occur at relatively low values of $m_A\sim130$--$250$ GeV. 
These low $m_A$ solutions were extremely difficult to generate with the top-down
approach, and indicate that they have a high degree of fine-tuning.\footnote{
The TW paper (Ref. \cite{tw}) remarks that there must be considerable fine-tuning as well
to reconcile $BF(b\to s\gamma )$ with Yukawa unification and the dark matter relic abundance.}
A scattering of DM-allowed dots occur with high $m_A$ values. These turn out to be
the $h$-resonance solutions as generated with the GSH boundary conditions in Sec. \ref{sec:h}.

\FIGURE[p]{
\epsfig{file=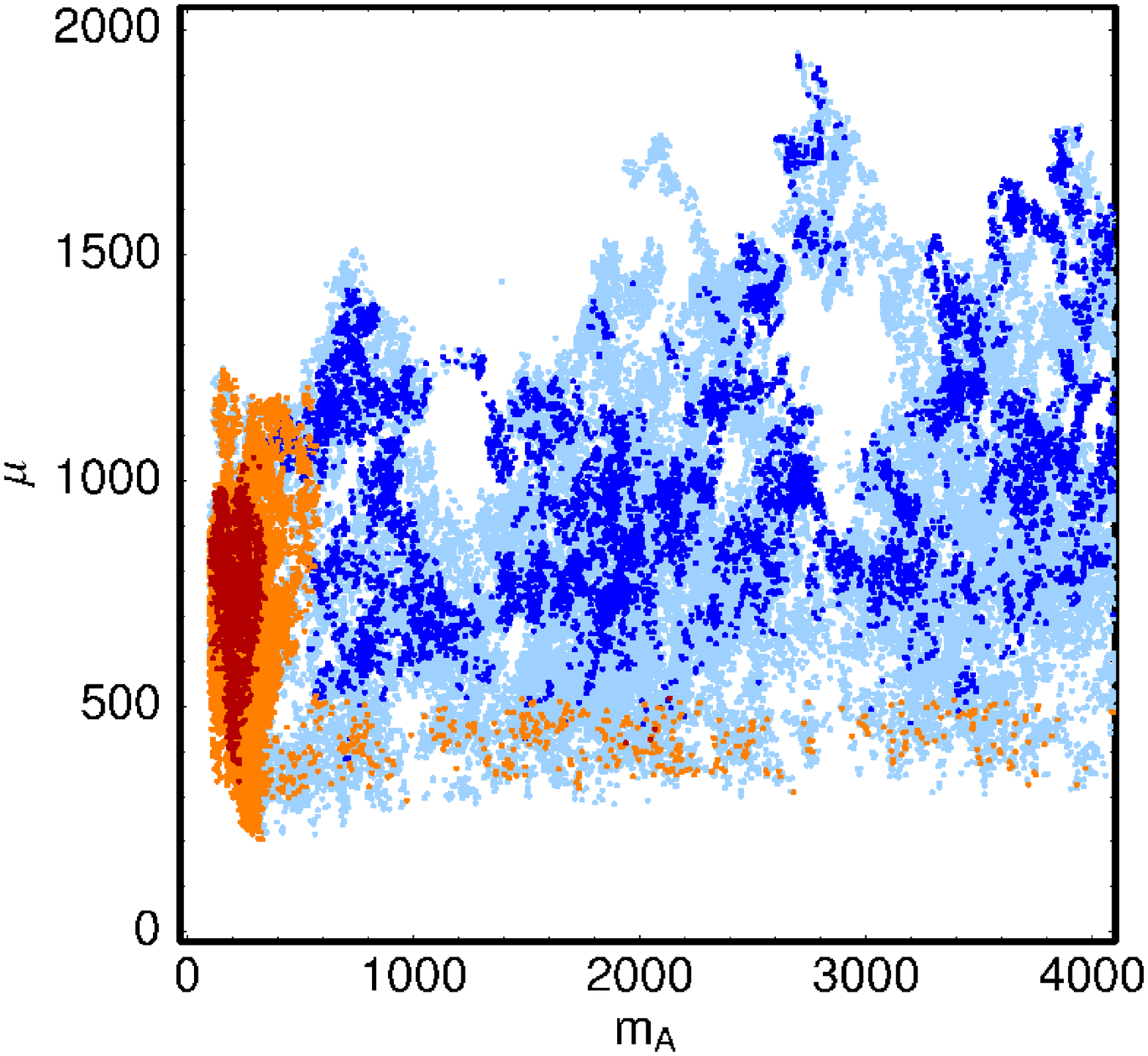,width=9cm} 
\caption{Plot of MCMC results using WSH boundary conditions 
in the $m_A$ vs. $\mu$ plane; the light-blue (dark-blue) points have $R<1.1\ (1.05)$, 
while for the orange (red) points $R<1.1\ (1.05)$ and $\Omega_{\tz_1}h^2<0.136$.}
\label{fig:pma0muBDR}}

This is seen more clearly by plotting in the $m_h-2m_{\tz_1}$~vs.~$m_A-2m_{\tz_1}$
plane in Fig.~\ref{fig:hAvsdiffBDR}. Here we see a narrow strip at
$m_h-2m_{\tz_1}=0$ corresponding to $h$-resonance annihilation solutions, while we
also have a wider band of solutions at $m_A-2m_{\tz_1}=0$, which indicate
neutralino annihilation through the $A$-resonance. The width of the latter
band is due to the fact that the $A$ width can be quite wide-- typically a few GeV,
while the $h$-width is much narrower, of order 50 MeV. 

\FIGURE[p]{
\epsfig{file=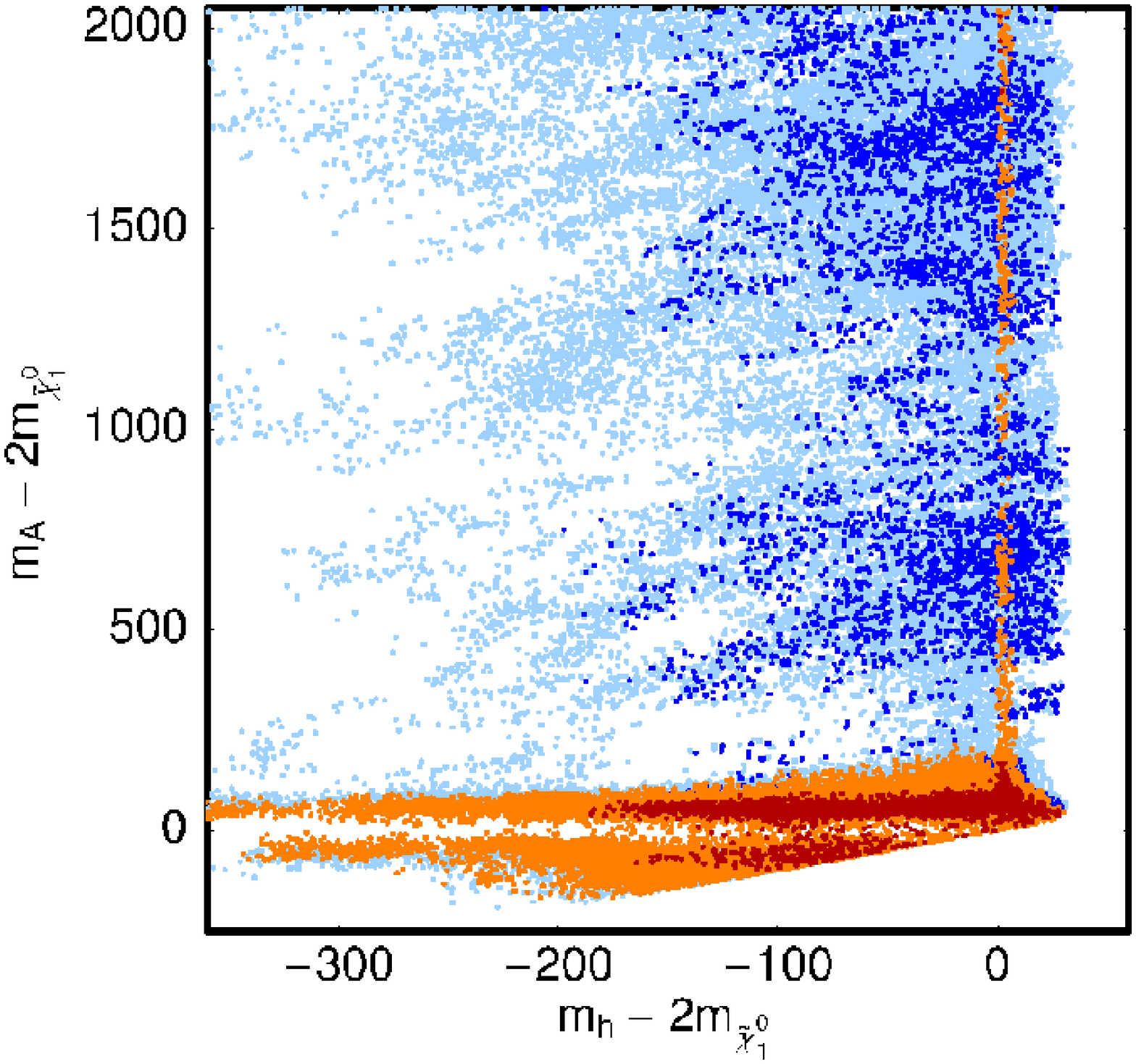,width=9cm} 
\caption{Plot of MCMC results using WSH boundary conditions 
in the $m_h-2m_{\tz_1}$ vs. $m_A-2m_{\tz_1}$ plane; the light-blue (dark-blue) points have $R<1.1\ (1.05)$, 
while for the orange (red) points $R<1.1\ (1.05)$ and $\Omega_{\tz_1}h^2<0.136$.}
\label{fig:hAvsdiffBDR}}

The $A$-resonance solutions occur at $\tan\beta\sim 50$ and relatively low $m_A$
values. This can signal dangerously high branching fractions for $B_s\to\mu^+\mu^- $
decay~\cite{Bstomumu} since the branching fraction goes like $\tan^6\beta/m_A^4$. We plot the
$BF(B_s\to\mu^+\mu^- )$~vs.~$m_h$ in Fig.~\ref{fig:pmh0bsmmBDR}. The 
recent experimental limit from the CDF collaboration is that $BF(B_s\to\mu^+\mu^- )<
5.8\times 10^{-8}$~\cite{cdflim}. Thus, the entire band of $A$-resonance annihilation 
solutions becomes excluded! The smattering of DM-allowed dots below the CDF limit
all occur with DM annihilation via the  $h$-resonance.

\FIGURE[p]{
\epsfig{file=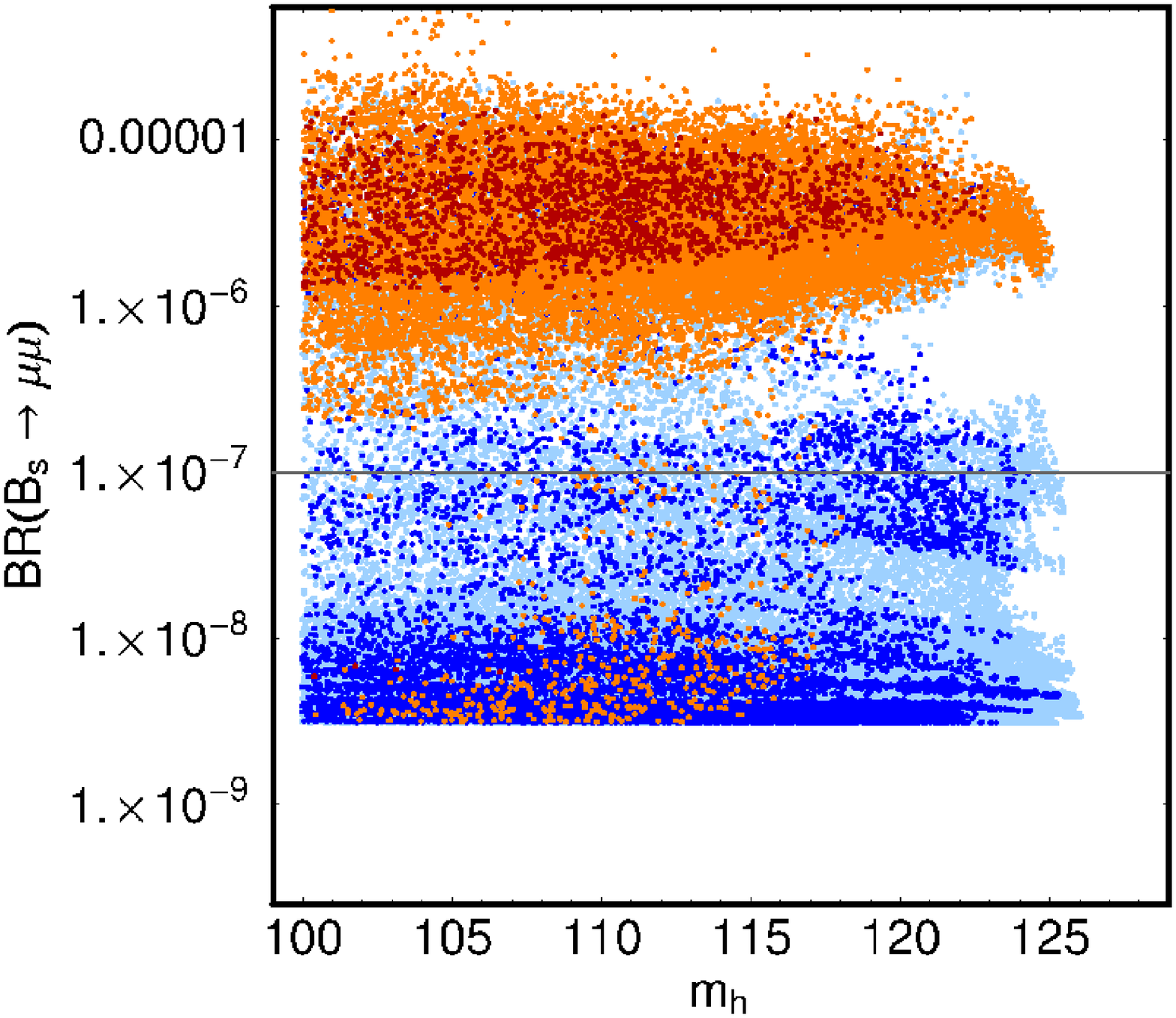,width=9cm} 
\caption{Plot of MCMC results using WSH boundary conditions 
in the $m_h$ vs. $BF(B_s\to\mu^+\mu^- )$ plane; the light-blue (dark-blue) points have $R<1.1\ (1.05)$, 
while for the orange (red) points $R<1.1\ (1.05)$ and $\Omega_{\tz_1}h^2<0.136$.}
\label{fig:pmh0bsmmBDR}}

In case these $A$-resonance solutions are somehow allowed-- say by additional
flavor-violating soft terms-- we plot the solutions in the $m_{\tg}$~vs.~$m_{\tz_1}$
plane in Fig.~\ref{fig:pmgmne1BDR}. Here, we see a much larger range of $m_{\tg}$
and $m_{\tz_1}$ values are DM-allowed than in the GSH solutions, with
$m_{\tg}$ extending up to 1500 GeV.

\FIGURE[p]{
\epsfig{file=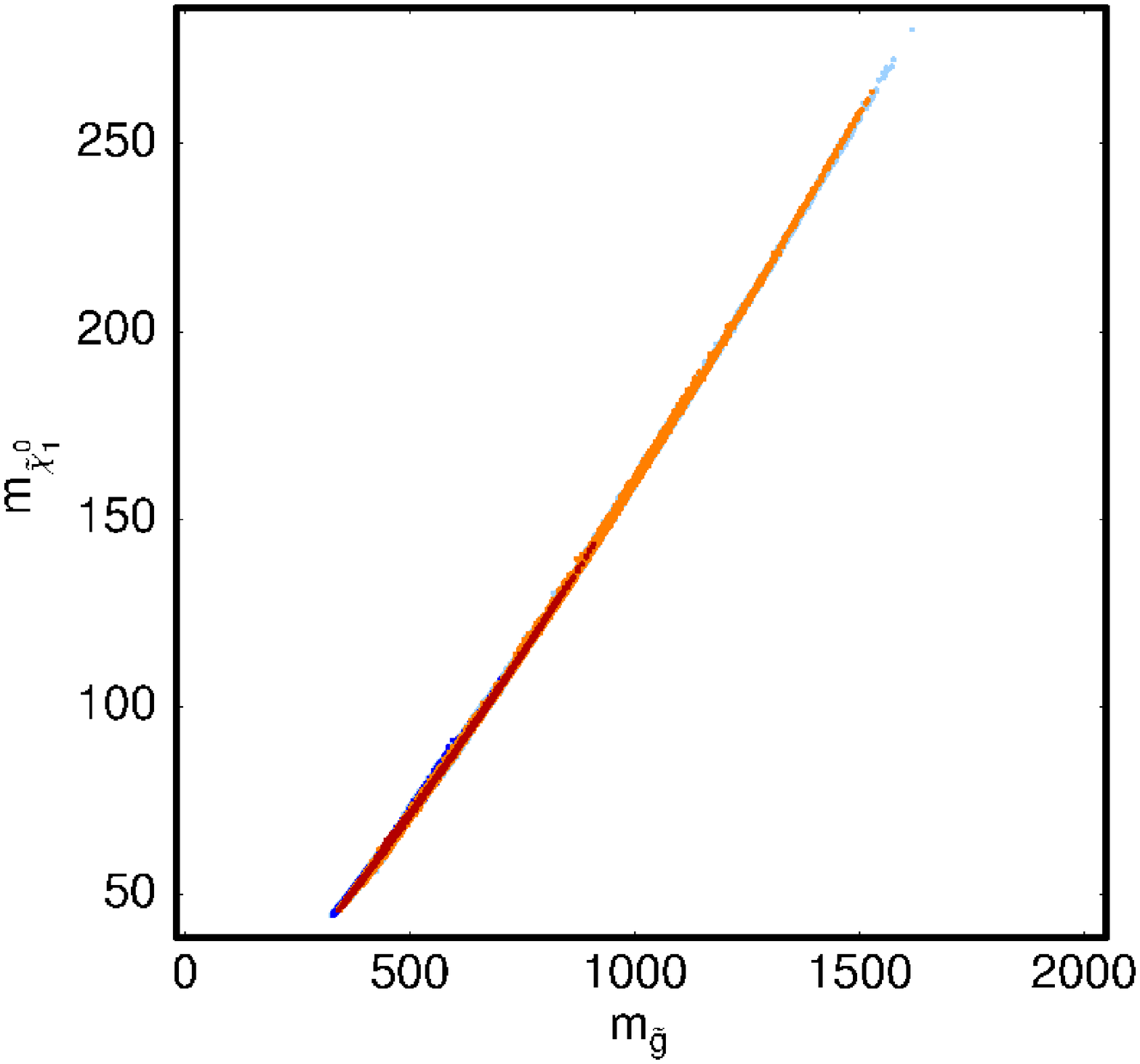,width=9cm} 
\caption{Plot of MCMC results using WSH boundary conditions 
in the $m_{\tg}$ vs. $m_{\tz_1}$ plane; the light-blue (dark-blue) points have $R<1.1\ (1.05)$, 
while for the orange (red) points $R<1.1\ (1.05)$ and $\Omega_{\tz_1}h^2<0.136$.}
\label{fig:pmgmne1BDR}}

If the DM-allowed GSH solutions are able to avoid the $BF(B_s\to\mu^+\mu^- )$
constraint, then the values of $m_A$ and $m_{\tz_2}-m_{\tz_1}$ will be correlated, 
and the latter quantity will be measureable if the mass gap  $m_{\tz_2}-m_{\tz_1}<M_Z$.
The predicted correlation is shown in Fig.~\ref{fig:pmne2Mmne1ma0BDR}.
In this case, the mass gap $m_{\tz_2}-m_{\tz_1}$ runs far beyond $M_Z$. 

\FIGURE[p]{
\epsfig{file=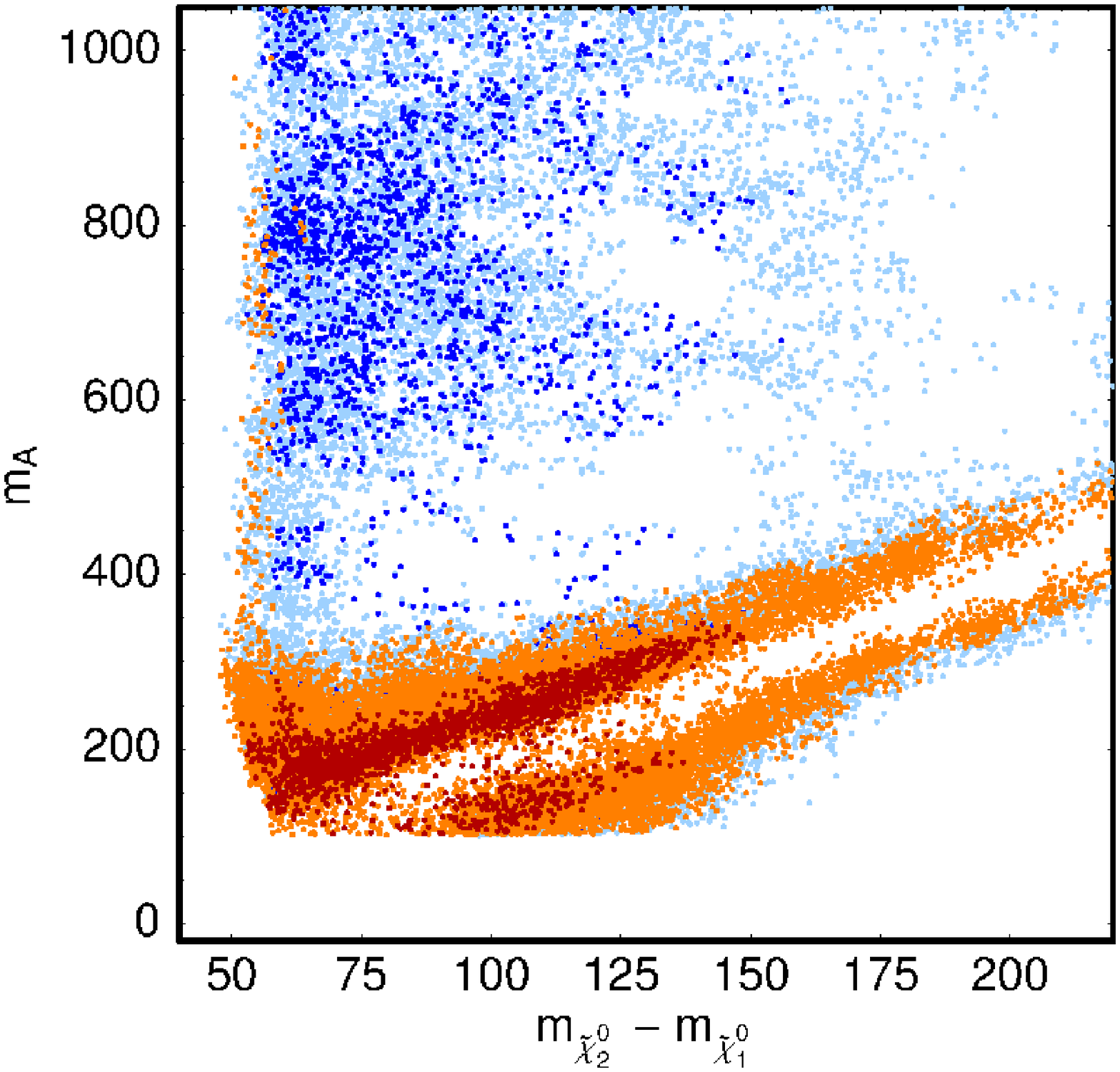,width=9cm} 
\caption{Plot of MCMC results using WSH boundary conditions 
in the $m_{\tz_2}-m_{\tz_1}$ vs. $m_A$ plane; the light-blue (dark-blue) points have $R<1.1\ (1.05)$, 
while for the orange (red) points $R<1.1\ (1.05)$ and $\Omega_{\tz_1}h^2<0.136$.}
\label{fig:pmne2Mmne1ma0BDR}}

We present a benchmark point with low $m_A$ and $R\simeq 1$ as point E in Table \ref{tab:bm}.
While the point is DM-allowed, it also violates the CDF bound on $BF(B_S\to \mu^+\mu^- )$.

At this point, it is useful to compare the Isajet SUSY spectral solutions
to those generated by Dermisek {\it et al.} in Ref.~\cite{drrr1} and \cite{drrr2}.
In Fig.~\ref{fig:bdr}, we plot the Isajet 7.75 solutions in the $m_{1/2}$~vs.~$\mu$
plane for $m_{16}=3$ TeV, $m_{10}/m_{16}=1.3$, $A_0/m_{16}=-1.85$, $\tan\beta =50.9$
and $\Delta m_H^2=0.14$, with $m_A=500$ GeV: {\it i.e.} corresponding
closely to Fig. 1 of \cite{drrr1}. We plot contours of $R$ from 1.15 to 1.3. 
Also, the green-shaded
regions give the WMAP-measured relic density, while white-shaded regions give 
$\Omega_{\tz_1}h^2<0.095$, and pink-shaded regions give $\Omega_{\tz_1}h^2>0.13$
 (as in Dermisek {\it et al.}). The
LEP2 constraint on $m_{\tw_1}$ is indicated by the solid contour at low 
$m_{1/2}$ and low $\mu$. We see qualitatively the same shape to the DM-allowed 
regions as generated by Dermisek {\it et al.}: the thick green regions are
DM-allowed either by $A$-resonance annihilation at large $\mu$, or by mixed higgsino
DM 

\FIGURE[p]{
\epsfig{file=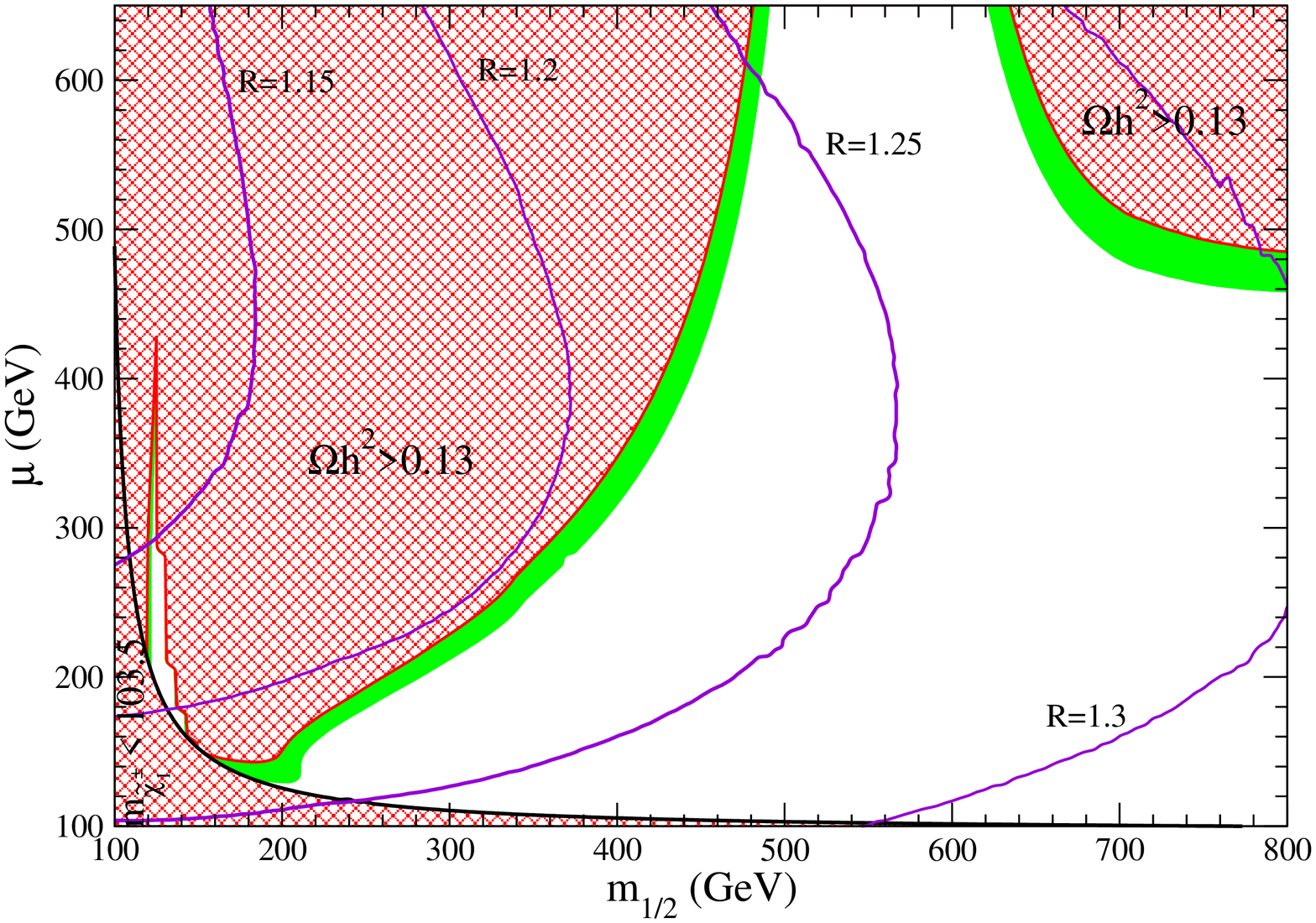,width=10cm} 
\caption{Contours of $R$ and DM-allowed regions in the 
$m_{1/2}$~vs.~$\mu$ parameter space for
$m_{16}=3$ TeV, $m_{10}/m_{16}=1.3$, $A_0/m_{16}=-1.85$, 
$\Delta m_H=0.14$, $\tan\beta =50.9$, $m_A=500$ GeV and
$m_t=173.9$ GeV, as in Dermisek {\it et al.}, but using
Isajet 7.75 for mass spectra generation.}
\label{fig:bdr}}

\clearpage

\noindent
annihilation at low $\mu$. There is also a light Higgs $h$-resonance solution at
$m_{1/2}\sim 120$ GeV. 

A notable feature of Fig.~\ref{fig:bdr} is that over much of the DM-allowed region, the Yukawa unification
has $R>1.2$.\footnote{Note that although the general features in Fig. \ref{fig:bdr} 
here and Fig. 1 of \cite{drrr1} are similar, the latter results were obtained 
in a top-down fit to low energy obervables assuming exact Yukawa unification, which is a different 
approach then the one followed here. Moreover, there are several important 
differences in the level of sophistication of the spectrum computations 
between Ref. \cite{bdr,drrr1,drrr2} and the study presented here. For instance, 
Ref. \cite{bdr,drrr1,drrr2} has only 
1-loop RGE running of the SUSY-breaking parameters, takes sparticle masses
to be running masses at scale $Q=M_Z$; ISAJET 7.75 
applies full 2-loop running plus 1-loop threshold corrections.
}
 As we move to larger $\mu$ values and lower $m_{1/2}$ values, the 
Yukawa unification gets better and better. Most of the region with $R<1.15$ is DM-forbidden, save for the
upper part of the light $h$-resonance solution. In fact, now we can see why our  compromise 
solution (point D) works and why it is so hard to find using a top-down approach: only the very narrow upper
tip is both DM-allowed, and has a low $R$ value.

\section{Yukawa-unified benchmark scenarios and LHC signatures}
\label{sec:lhc}

We have assembled in Table \ref{tab:bm} five Yukawa-unified benchmark scenarios that yield the correct
relic abundance of dark matter in five different ways. With the LHC turn-on being imminent, it is
fruitful to examine what each of these five scenarios implies for new physics signatures.

At the bottom of Table \ref{tab:bm} we list $\Omega_{\tz_1}h^2$, $BF(b\to s\gamma )$, 
$BF(B_s\to\mu^+\mu^- )$, $\Delta a_\mu$ and spin-independent 
neutralino-proton direct DM detection cross section $\sigma (\tz_1 p)$. 
For the first four of these numbers, we list output from IsaReD/Isatools (upper) and 
micrOMEGAs (lower). 
While the results for the low-energy constraints agree fairly well, there is almost a factor of 2 
difference in the relic density when the neutralino dominantly annihilates through $h$ or $A$ 
exchange (points A, D, E). This is due to differences in the treatment of the Higgs resonance. 
For example, IsaReD in Isajet 7.75 uses Yukawa couplings evaluated at scale 
$Q=\sqrt{m_{\tst_L}m_{\tst_R}}$ for annihilation through the $A$ resonance and for evaluation 
of the heavy Higgs widths, while micrOMEGAs uses an effective Lagrangian approach and  
$Q=2m_{\tz_1}$.\footnote{A complete discussion of the details of the calculations in the two 
programs is beyond the scope of this paper; we refer the interested reader to the respective manuals.}

\subsection*{Point A}

Point A of Table \ref{tab:bm} is a generic Yukawa-unified model with first and second generation 
scalar masses $\sim 9$ TeV, so they essentially decouple from LHC physics. Third generation and 
heavy Higgs scalars have masses at the 2--3 TeV level, while the lightest charginos, neutralinos and gluinos
all have masses in the range 100--400 GeV. Since $\Omega_{\tz_1}h^2\sim 400$, we postulate that 
the neutralino $\tz_1$ is in fact an NLSP, decaying to $\ta\gamma$ with a lifetime of order
0.03 seconds. In this case, the mean decay distance of a $\tz_1$ will be of order 
$10^4$ km. Thus, the $\tz_1$ will still escape the LHC detectors, leading to missing energy signatures
(although it is conceivable some may decay occassionally within the detector).

The LHC SUSY events will consist of a hard and soft component~\cite{gabe}. 
The hard component comes from pair production
of $\sim 400$ GeV gluinos. The gluinos decay via 3-body modes dominantly via $\tg\to tb\tw_1$, $b\bar{b}\tz_1$
and especially $b\bar{b}\tz_2$~\cite{ltanb}. 
The $\tg\tg$ production cross section is of order $10^5$ fb at LHC, so we might
expect $10^7$ gluino pair events per 100 fb$^{-1}$ of integrated luminosity. After cascade decays, we expect
an assortment of events with high jet and $b$-jet multiplicity, plus an assortment of isolated leptons.
The $\tz_2\to\tz_1 e\bar{e}$ branching fraction is at 2.2\% , which should be enough to reconstruct
the dilepton mass edge at $m_{\tz_2}-m_{\tz_1}\simeq 73$ GeV. 
Correct pairing of $b$-jets and/or
$b$-jets with isolated leptons, plus the total event rate, 
should allow for an extraction of the gluino mass.

The soft component of signal will come from $\tilde\chi_1^+\tilde\chi_1^-$, $\tw_1\tz_2$ 
and $\tw_1\tz_1$ production.
These events will be followed by 3-body decays to various final states, but since the
visible components of the signal are much softer than that from gluino pair production, these
events will be harder to see above SM background levels. With judicious cuts, 
the soft component might also
be visible at some level ({\it e.g.} $\tw_1\tz_2\to 3\ell +\eslt$)~\cite{lhc3l}.

%
\begin{table}
\begin{tabular}{lccccc}
\hline
parameter & A & B & C & D & E \\
\hline
$m_{16}$ & 9202.9 & 9202.9 & 5018.8 & 2976.5 & 5877.3 \\
$m_{1/2}$ & 62.5 & 62.5 & 160 & 107.0 & 113.6 \\
$A_0$ & $-$19964.5 & $-$19964.5 & $-$10624.2 & $-$6060.3 & $-$12052.6 \\
$m_{10}$ & 10966.1 & 10966.1 & 6082.1 & 3787.9 & --- \\
$\tan\beta$ & 49.1 & 49.1 & 47.8 & 49.05 & 47.4 \\
$M_D$ & 3504.4 & 3504.4 & 1530.1 & 1020.8 & ---  \\
$M_1$ & --- & 195 & --- & --- & --- \\
$m_{16}(1,2)$ & --- & ---  & 603.8 & --- & --- \\
$f_t$ & 0.51 & 0.51 & 0.49 & 0.48 & 0.49  \\
$f_b$ & 0.51 & 0.51 & 0.41 & 0.47 & 0.49  \\
$f_\tau$ & 0.52 & 0.52 & 0.47 & 0.52 & 0.49 \\
$\mu$ & 4179.8 & 4186.3 & 1882.6 & 331.0 & 865.3 \\
$m_{\tg}$   & 395.6 & 395.4 & 495.5 & 387.7 & 466.6 \\
$m_{\tu_L}$ & 9185.4 & 9185.4 & 622.1 & 2970.8 & 5863.0 \\
$m_{\tu_R}$ & 9104.1 & 9104.2 & 98.3 & 2951.4 & 5819.2 \\
$m_{\tst_1}$& 2315.1 & 2310.5 & 1048.4 & 434.5 & 944.7  \\
$m_{\tb_1}$ & 2723.1 & 2714.9 & 1894.0 & 849.3 & 1452.7 \\
$m_{\te_L}$ & 9131.9 & 9132.0 & 311.9 & 2955.8 & 5833.6 \\
$m_{\te_R}$ & 9323.7 & 9323.9 & 891.8 & 3009.0 & 5945.8  \\
$m_{\tw_1}$ & 128.8 & 128.8 & 165.7 & 105.7 & 141.3 \\
$m_{\tz_2}$ & 128.6 & 128.1 & 165.1 & 105.1 & 140.9 \\ 
$m_{\tz_1}$ & 55.6 &  115.9 & 80.2 & 52.6  & 65.7 \\ 
$m_A$       & 3273.6 &  3266.0 & 1939.9 & 776.8 & 177.8 \\
$m_h$       & 125.4 &  125.4 & 123.2 & 111.1 & 113.4  \\ \hline
$\Omega_{\tz_1}h^2$& $423\atop 220$  & $0.09\atop 0.08$ & $0.11\atop 0.11$ & $0.10\atop 0.06$ & $0.15\atop 0.08$  \\
$BF(b\to s\gamma)$ & $3.0\times 10^{-4}\atop 3.3\times 10^{-4}$ & $3.0\times 10^{-4}\atop 3.3\times 10^{-4}$ & 
$6.2\times 10^{-4}\atop 3.7\times 10^{-4}$ & $1.9\times 10^{-4}\atop 4.0\times 10^{-4}$ 
& $2.5\times 10^{-4}\atop 2.2\times 10^{-4}$ \\
$\Delta a_\mu    $ & $5.0 \times  10^{-12}\atop 5.1\times 10^{-12}$ & $5.0 \times  10^{-12}\atop 5.0\times 10^{-12}$ 
& $3.0 \times  10^{-10}\atop 2.8\times 10^{-10}$ & $2.2 \times  10^{-10}\atop 2.2\times 10^{-10}$ 
& $4.1\times 10^{-11}\atop 4.1\times 10^{-11}$ \\
$BF(B_s\to\mu^+\mu^- )$ & $5.0\times 10^{-9}\atop 4.4\times 10^{-9}$ & $5.0\times 10^{-9}\atop 4.4\times 10^{-9}$ 
& $11.8\times 10^{-9}\atop 6.9\times 10^{-9}$ & $5.8\times 10^{-8}\atop 6.2\times 10^{-8}$ & 
$2.0\times 10^{-5}\atop 2.0\times 10^{-5}$  \\
$\sigma_{sc} (\tz_1p )\ [{\rm pb}]$ 
  & $1.3\times 10^{-15}$ 
  & $1.9\times 10^{-17}$ 
  & $1.5\times 10^{-6}$ 
  & $2.7\times 10^{-9}$ 
  & $5.3\times 10^{-8}$ \\
\hline
\end{tabular}
\caption{Masses and parameters in~GeV units
for five benchmark Yukawa unified points using Isajet 7.75
and $m_t=171.0$ GeV.
The upper entry for the $\Omega_{\tz_1}h^2$ etc. come from IsaReD/Isatools, while the lower
entry comes from micrOMEGAs; $\sigma (\tz_1p )$ is computed with Isatools.}
\label{tab:bm}
\end{table}

\subsection*{Point B}

Point B is the same as point A, except that in this case the gaugino mass $M_1$ has been raised to
195 GeV so that the $\tw_1 -\tz_1$ mass gap shrinks to only 13 GeV. Since $\mu$ is quite large, 
the $\tz_1$ remains nearly pure bino-like, but the relic density problem is solved via
bino-wino co-annihilation. This case will again give a hard component to the LHC new physics signal from
gluino pair production, but this time the $m(\ell^+\ell^- )$ distribution will have an edge only at 13 
GeV. When compared to any gluino mass reconstructions, this would indicate a violation of
gaugino mass unification at the GUT scale. 
In addition, the small $\tz_2-\tz_1$ mass gap suppresses 3-body decays such as $\tz_2\to \tz_1 q\bar{q}$ and
$\tz_1\ell\bar{\ell}$ relative to any kinematically-allowed 2-body decays such as the loop-induced
process $\tz_2\to\tz_1\gamma$~\cite{z2z1g}.
Thus, the radiative $\tz_2$ decay to photon $\tz_2\to\tz_1\gamma$ can become large~\cite{bwca}:
in this case, it reaches 10\%. 
The final state $\gamma$ will be somewhat soft if the $\tz_2$ is at low velocity. 
But if $\tz_2$ is moving fast as a result of production from cascade decays, then 
hard, isolated photons should occasionally be present in the SUSY collider events.

\subsection*{Point C} 

The point C parameters are listed in Table \ref{tab:bm}. In this case, 
$m_{16}(1,2)$ has been lowered far below $m_{16}(3)$ 
so that the first two generations of scalars are degenerate, but with a 
lower mass than third generation scalars. The Higgs mass splitting
leads to a large RGE $S$-term, which drives $\tu_R$ and $\tc_R$ to very low masses $\simeq 98.3$ GeV. 
This is at the edge of LEP2 exclusion.
The relic density problem is solved because $\tz_1\tz_1\to q\bar{q}$ via $\tq_R$ exchange
and neutralino-squark co-annihilation act to reduce the relic density. 
The cross section for production of two flavors of extremely light squarks
is extremely large at LHC. Normally one would expect characteristic dijet$+\eslt$ events since 
$\tq_R\to q\tz_1$. However, in this case the mass gap $m_{\tu_R}-m_{\tz_1}\sim 18$ GeV, so both the
jets and $\eslt$ will be very soft.
Gluinos and other squarks will also be produced at large rates, although the $\tg\to  u\tu_R,\ c\tc_R$
decays are dominant.
While left-squarks may decay with large rates to $\tz_2$ and $\tw_1$, we note that $\tz_2\to u\tu_R$
and $c\tc_R$ is also large, leading again to relatively soft jet activity.
In spite of the soft jet activity, the scenario should be easily seen at LHC, since $\tq_L\to q'\tw_1$
occurs at a large rate, and $\tw_1\to e\nu_e\tz_1$ occurs at 43\% branching fraction
(enhanced by the relatively light left-sleptons). 
This can lead to a large same-sign (SS) dilepton rate from $pp\to\tu_L\tu_L$ production, 
along with a large asymmetry in $++$ SS dileptons over $--$ SS dileptons (which occur from
$\td_L\td_L$ production).
This scenario may also be subject to exclusion by analysis of Fermilab Tevatron data.
We further note that point C is naively excluded by direct dark matter search limits. 
These latter limits depend on an assumed standard local relic density mass and velocity distribution,
so that the limits can be avoided if one postulates that we live in a local underdensity of 
dark matter.

\subsection*{Point D}

Point D is an example of a compromise solution, where we allow $m_{16}$ as low as 3 TeV at some expense
to Yukawa unification (here, Yukawa unification is good to only $\sim 10\%$) in order to allow for neutralino
annihilation through the light Higgs $h$-resonance (neutralinos can still annihilate through the
light $h$ resonance for higher $m_{16}$ values; it is just that the relic density can't be pushed as low
as $\Omega_{\tz_1}h^2\sim 0.1$). This scenario is extremely predictive, with gluinos around 350--450
GeV, so again we expect LHC events to be dominated by gluino pair production. 
As in the case of point A, the $\tg\tg$ events will be followed by 3-body decays to $b$-jet rich final states.
A dilepton mass edge at $m_{\tz_2}-m_{\tz_1}\simeq 53$ GeV should be visible since $\tz_2\to\tz_1 e^+e^-$ at 
3.3\% branching fraction.
The $\tilde t_1$ weighs only 434 GeV in this case, and $\tilde b_1$ is at 849 GeV, so it may be possible 
to detect some third generation squark pair production events. 
The top squark decays to $b\tw_1$ with a 50\% branching fraction, and also has significant
branching fractions to $t\tz_1$, $t\tz_2$ and $b\tw_2$ final states. 
The $\tb_1$ dominantly decays to $b\tg$ and $W\tst_1$ final states.
Moreover, the heavy Higgs bosons $A^0$, $H^0$ and $H^\pm$ have masses around 780 GeV and 
should be detectable at LHC~\cite{tdr}.

\subsection*{Point E}

Point E is a Yukawa-unified solution that solves the DM abundance problem via
neutralino annihilation through a 178 GeV pseudoscalar $A$ resonance. The combination of
light $A$ and large $\tan\beta$ leads to a branching fraction $B_s\to\mu^+\mu^-$ which is excluded
by recent CDF analyses. If we are allowed to somehow ignore this (possibly via other flavor-violating 
interactions), then the scenario would be at the edge of observability via Tevatron searches
for $A,\ H\to\tau^+\tau^-$ and $b\bar{b}$, which at present exclude $m_A\alt 170$ GeV~\cite{tevAexclusion}.
The LHC (and possibly soon also the Tevatron) would easily see the rather light spectrum
of Higgs bosons. Gluinos can be somewhat heavier in this case compared to points A and D, 
ranging to over a TeV.
However, in point E as listed, with a 467 GeV gluino, the gluino pair production signatures
will be rather similar to those of point A: rich in $b$-jets, with a visible dilepton mass edge 
at 75 GeV.

\section{Summary and conclusions}\label{sec:conclude}

In this paper, we have presented a number of new results.
\begin{enumerate}
\item First, we verified former results presented in Ref.~\cite{abbbft} that Yukawa unified models
can be generated with updated Isajet spectra code and an updated value of the top quark mass
$m_t=171$ GeV. Using both random scans and the more efficient MCMC scans, we find that models
with excellent Yukawa coupling unification can be generated in the HS model if scalar masses are
in the multi-TeV range, while gaugino masses are quite light, and the $\tw_1$ is slightly above the 
current LEP2 limit. The models require the Bagger {\it et al.} boundary conditions if $\mu >0$
such that $A_0^2=2m_{10}^2=4 m_{16}^2$, and $A_0<0$ in our convention. The spectra generated is
characterized by three mass scales: multi-TeV first and second generation matter scalars, TeV scale
third generation and Higgs scalars and 100--200 GeV light charginos and gluinos of order
350--450 GeV.
The relic density is typically 30--30,000 times above the WMAP measured value. As a solution, 
we propose {\it i}). hypothesizing an unstable neutralino $\tz_1$ which decays to axino plus photon,
{\it ii}). raising the GUT scale gaugino mass $M_1$ so that bino-wino co-annihilation reduces the relic density
or {\it iii}). lowering the first/second generation scalar masses relative to the third so that
neutralinos can annihilate via light $\tq_R$ exchange and neutralino-squark co-annihilation.
We regard the first of these solutions as the most attractive, and the third is actually susceptable
to possible exclusion by analyses of Fermilab Tevatron signals in the case of just two light squarks.
\item Using an MCMC analysis, we find a new class of solutions with $m_{16}\sim 3$ TeV, 
where neutralinos annihilate through the light higgs $h$ resonance. This low a value of $m_{16}$
typically leads to Yukawa unification at the 5--10\% level at best.
\item We find we are able to generate solutions with low $\mu$ and low $m_A$ as did the BDR group.
The solutions generated by the Isajet code with low $\mu$, low $m_A$ and $m_{16}\sim 3$ TeV tend to have
Yukawa unification in the 20\% range or greater. We were able to generate a class of solutions
with excellent Yukawa unification and $m_{16}$ ranging up to 6 TeV, where the DM problem is solved
by neutralino annihilation through a $150$--$250$ GeV $A$ resonance. The combination of large $\tan\beta$
and low $m_A$ gives a $B_s\to\mu^+\mu^-$ branching fraction at levels beyond those allowed
by the CDF collaboration.
\end{enumerate}

We also present a Table of five benchmark solutions suitable for event generation, and for examination
of collider signals expected at the LHC from DM-allowed Yukawa-unified SUSY models. 
Based on this work, we are able to make several predictions, if the Yukawa-unified MSSM is the
correct effective field theory between $M_{GUT}$ and $M_{weak}$.
We would expect the following:
\begin{itemize}
\item New physics events at the CERN LHC to be dominated by gluino pair production with 
$m_{\tg}\sim 350$--$450$ GeV. Since $\tan\beta$ is large, the final states are rich in $b$-jets, 
and the OS/SF isolated dilepton invariant mass distribution should have a visible edge
at $m_{\tz_2}-m_{\tz_1}\sim 50$--$75$ GeV because the $\tz_2$ always decays via 3-body modes.
Squarks and sleptons are likely to be very heavy, and may decouple from LHC physics signatures.
\item We would predict in this scenario that the $(g-2)_\mu$ anomaly is false, since in Yukawa-unified
SUSY models with large $m_{16}$, the SUSY contribution to the muon QED vertex is always
highly suppressed. 
\item While SUSY should be easily visible at the LHC for Yukawa unified models, 
we would predict a dearth of direct and indirect dark matter detection signals. This is because
the typically large values of $\mu$ and scalar masses tend to suppress such signals. However, in the 
CDF-excluded case of point E, the direct and indirect DM signals may be observable.
Points C and D also have low but observable direct DM detection rates, since scalars are not too heavy.
\end{itemize}

\acknowledgments

We thank G. Belanger, A. Belyaev, A. Pukhov and X. Tata for useful comments and discussion,
and A. Pukhov for finding a bug in the IsaReD relic density subroutine.
This project was initiated during the CERN-TH institute on {\it LHC--Cosmology Interplay},
where three of us (HB, SK and SS) participated. The hospitality of the CERN TH division is 
gratefully acknowledged. SS additionally acknowledges the support by T\"UB\'ITAK grant 106T457 (TBAG-HD 
190), enabling her visit to CERN.  This research was supported in part by the U.S. Department of Energy
grant numbers DE-FG02-97ER41022.

%

\end{document}